\documentclass[fleqn,usenatbib]{mnras}

\usepackage{newtxtext,newtxmath}

\usepackage[T1]{fontenc}

\DeclareRobustCommand{\VAN}[3]{#2}
\let\VANthebibliography\thebibliography
\def\thebibliography{\DeclareRobustCommand{\VAN}[3]{##3}\VANthebibliography}

\usepackage{graphicx}	
\usepackage{amsmath}	
\usepackage{caption}
\usepackage{subcaption}
\usepackage{rotating}
\usepackage{tablefootnote}

\newcommand{\DTD}{\texttt{Dust2Dust}}
\defcitealias{Dovekie}{Dovekie}
\defcitealias{Grayling24}{G24}
\defcitealias{Thorp21}{T21}

\title[Stjörnumál]{A Simulation Based Inference Approach to Modelling of Type Ia Supernova Populations }

\author[Popovic et al.]{
B. Popovic$^1$\thanks{Email: B.A.Popovic@soton.ac.uk} , M. Grayling$^2$, M. O'Callaghan$^2$,  B. M. Boyd$^2$, K. Mandel$^2$, P. Wiseman$^1$, B. Carreres$^3$, 
\newauthor
N. Shiamtanis$^1$, D. Scolnic$^3$, E. Charleton$^1$, Y. Murakami$^4$
\\
$^1$ School of Physics and Astronomy, University of Southampton, Southampton, SO17 1BJ, UK \\
$^2$ Institute of Astronomy and Kavli Institute for Cosmology, University of Cambridge, Madingley Road, Cambridge, CB3 0HA, UK \\
$^3$ Department of Physics, Duke University, Durham, NC 27708, USA \\
$^4$ Department of Physics and Astronomy, Johns Hopkins University, Baltimore, MD 21218, USA
}

\date{Accepted XXX. Received YYY; in original form ZZZ}

\pubyear{2026}

\begin{document}
\label{firstpage}
\pagerange{\pageref{firstpage}--\pageref{lastpage}}
\maketitle

\begin{abstract}
Type Ia Supernovae (SNe Ia) are prominent cosmological probes, utilising a standardisation process to reduce their observed scatter to $\sim0.15$ mag. A growing number of models seek to explain the remaining scatter, based on a diversity of dust properties and possible connections to the progenitor systems. Inference of new models has been limited due to cost of simulations and complexity. 
We present Stjörnumál, a simulation based inference pipeline to infer intrinsic and extrinsic parameters of SNe Ia. Stjörnumál provides fast and accurate posterior inference via Neural Posterior Estimation, integrated model comparison with Neural Ratio Estimation, and overall significant speed and quality-of-life upgrades. We fit the Dark Energy Survey (DES) 5-year SN sample, finding good agreement with previously-published dust model parameters for DES5YR. We test 7 models of SN Ia behaviour, finding more data is needed to break degeneracies between $R_V$ models, but sufficient to evidence ($\log(10)\textrm{Bayes Factor} = +1.9$, $f_{\rm mix} = 0.8$) against two populations of SNe Ia at high-redshift. We employ a combination of frequentist $\chi^2$ metrics and Bayesian model comparison to make model determinations, finding neither are sufficient alone to properly compare models. For our nominal model, we find a smaller $\Delta R_V = 0.8$ than previous SALT-based attempts. We test model consistency against assumed cosmology, and find our results are robust to $|\Delta w| < 0.10$. The code is publicly available at https://github.com/bap37/Stjornumal, and presents an opportunity to flexibly and rapidly test potential models of SNe Ia scatter in a common framework. 
\end{abstract}




\section{Introduction}\label{sec:Intro}

Type Ia supernovae (SNe Ia) were instrumental in the discovery of the accelerating expansion of the universe by \cite{Riess98,Perlmutter99}. In the intervening decades, the number and attendant precision of following cosmology analyses with SNe Ia have only increased, from $\sim50$ SNe Ia to thousands today. Due to their luminosity ($\sim -19$ mag at peak) and low scatter ($\sim 0.15$ mags after standardisation), they remain crucial for probing cosmology to this day \citep{Brout22,Rubin25, DES5YR, Popovic26}. 

This success relies on the standardisation of SNe Ia: bluer SNe Ia are typically brighter \citep{Phillips93}, and SNe with a longer light curve width are also brighter. These correlations have been incorporated into a linear standardisation relationship \citep{Tripp98} that combine these features with the peak brightness. In the modern era, this standardisation is typically done after light curve fitting, performed with software such as SALT \citep{Guy07,Guy10,Taylor21,Kenworthy21}, or, more recently, BayeSN \citep{Mandel09,Mandel11,Thorp21,Grayling24}.

However, the construction of surveys of SNe Ia are beset by selection effects, most notably Malmquist bias, and deeper understanding of the intrinsic scatter of SNe Ia. Further complicating  the picture is the empirical nature of standardisation: the `true' distance-magnitude relationship may not be entirely described by the standardisation procedure.
In SN Ia samples, this will bias the population-level distributions towards the aforementioned brighter and bluer objects. These selection effects are degenerate with cosmology, and as there are not always analytical solutions to these complex effects, their interplay with astrophysical processes remain one of the largest sources of systematic uncertainty in cosmology to this day \citep{DES5YR, Dhawan25, Wiseman26}.

Selection effects and the nature of SN Ia intrinsic scatter are mitigated via two main methods in modern cosmology analyses. The first, UNITY \citep{Rubin25}, assumes that selection can be modelled as a simple analytical function (e.g. a sigmoid) that is inferred simultaneously with population-level parameters in the form of a Bayesian Hierarchical Model. This approach relies on the assumption that the selection effects \textit{can} be modelled with a simple analytic form. 

We focus more on the second approach, BEAMS with Bias Corrections (BBC, \citealp{Kessler16}). BBC uses realistic simulations generated by \texttt{SNANA} \citep{SNANA, Kessler19}, which model supernova populations with physical effects such as host galaxy association and contamination, dust extinction, and population-level differences while also including survey effects that mock observing strategies, cadence, filters, and sky and other sources of noise. The instrumental and survey aspects, inclusive of detection thresholds, survey changes, and gaps in coverage are incorporated through a combination of observing logs and filter measurements. Biases are then corrected for by comparing the difference between the true simulated parameters and the observed ones.

However, both methodologies entirely rely on a model of SN Ia scatter that replicates the data. Improved modelling of SNe Ia has had a long history, from initial dustless Spectral Energy Distribution (SED)-only approaches \citep{Guy10, Chotard11} to phenomenological approaches that seek to recreate the observed non-Gaussian behaviour \citep{Scolnic16,Popovic21a}. In the last several years, focus on the astrophysical modelling has only increased, leading to a cornucopia of new models that seek to explain the now-dominant systematic uncertainty. 

Of particular success in its implementation in cosmology analyses has been the dust model introduced by \cite{BS20}, further improved by \cite{Popovic22} which ascribes the variation observed in SNe Ia to a single population of SNe impacted by differing dust distributions. However, other works, such as \cite{Gonzalez-Gaitan21,Wojtak23,Rubin26}, suggest multiple populations of SNe Ia, and works such as \cite{Rigault18,Wiseman22,Popovic24b} suggest that dust is not the only component, and that differing progenitor channels may impact the observed behaviour of SNe Ia. 

The \texttt{Dust2Dust} code \citep{Popovic22} was originally designed to extract dust and intrinsic population information from samples of SNe Ia by modelling distributions of SALT parameters and their relationships with host galaxies and local environments, but had several limitations, most notably the long convergence time and inability to compare models in a statistically-principled way. An upcoming paper, Carreres et al. \textit{in prep.}, goes into more detail on the issues with \texttt{Dust2Dust}, including the impact on cosmology, alongside a novel Bayesian Hierarchical Model approach. Since the publishing of \texttt{Dust2Dust}, simulation based inference methods (SBI, \citealp{Cranmer20}) have undergone rapid improvement. Within SBI, simulations provide the otherwise intractable likelihoods that describe real data, using neural networks to translate the myriad simulations into understanding of the data. The neural networks can take many forms, such as neural density estimation \citep{Rezende15,Papamakarios19} or neural ratio estimation \citep{Hermans20}. SBI has been successful in astronomy and cosmology, including previous applications to SNe Ia with eye towards the posterior distribution and likelihood-to-evidence ratio \citep{Karchev24, Karchev26, Boyd24, Boyd26, OCallaghan25}. 

To ameliorate the issues of the original \texttt{Dust2Dust}, we replace the MCMC framework within \texttt{Dust2Dust} with an SBI system, and additionally include a neural ratio estimation function to compare and contrast models proposed within the literature. This new framework is called Stjörnumál. The layout of the paper is then as follows: Section \ref{sec:Data} provides an overview of the data and simulations. Section \ref{sec:Model} presents the lightcurve fitting and dust effects, followed by Section \ref{sec:Methodology} with an overview of the methodology employed. In Section \ref{sec:Models} we describe each of the SN Ia models we test, and explain the SBI pipeline in Section \ref{sec:SBI}. Finally, we show validation, results, and our conclusions in Sections \ref{sec:Validation}, \ref{sec:Results}, \ref{sec:Discussion}.

\section{Data and Simulations}\label{sec:Data}

\subsection{Data}\label{sec:Data:subsec:Data}

We use the homogeneous, high redshift sample of SNe Ia  provided by the DES collaboration \cite{DES5YR,DESCOSMO,Sanchez24}, with updated calibration by \cite{Popovic26}. This sample, `DES-Dovekie', ranges from $0.10 < z <1.13$. We do not use the complementary low redshift sample of $\sim200$ SNe from non-DES sources.

The DES-Dovekie data release includes the fitted SALT parameters and standardised brightnesses (Section \ref{sec:Model}), as well as bias corrections, redshifts, and other information. We do \textit{not} make use of the final cosmological distances in this analysis; instead we use the un-corrected distances that we reconstruct from the SALT parameters.

We relax the strict cosmology-cuts\footnote{Best laid-out in Table 4 of \cite{DES5YR}} so as to no longer constrain the tails of the $x_1$ and $c$ distributions:
\begin{itemize}
    \item ``Normal SNIa'' $(|x_1| < 3 ~\&~|c| < 0.3)$
    \item `Well constrained` $(\sigma_{x1} < 1, \sigma_{\rm t_{peak}} < 2)$
    \item Chauvenets criterion
    \item Valid bias correction
    \item Common subsample
\end{itemize}
and implement two new cuts: a valid host-galaxy mass, and a likelihood of being a type Ia SNe Ia of $P_{\rm Ia} > 0.5$\footnote{The strong bimodal nature of $P_{\rm Ia}$ in DES means that a cut of $P_{\rm Ia}$ removes most contaminants.}. These changes in data preparation leave us with 2201 likely SNe Ia. 


\subsection{Simulations}\label{sec:Data:subsec:Sims}

To create the training simulations, we use the \texttt{SNANA} simulation software from \cite{Kessler09,Kessler19}. For a given light curve model -- here SALT -- SNANA will generate a rest-frame SED model with selected model parameters, and applies a combination of galactic effects and cosmological effects before integrating the SED through specified filters, mimicking survey PSF, zero points, and source flux. Additional survey-specific criteria are then applied; for instance DES requires 2 detections across two different nights, before final selection criteria are applied to match observing efficacy. We make use of the DES simulations from \cite{DES5YR,Popovic26} using the SALT3.DOVEKIE model; more information on the specifics of DES simulations within \texttt{SNANA} can be found there.

When simulating, our nominal reference simulation is generated with a flat $\Lambda$CDM cosmology with $\Omega_M = 0.3, H_0 = 73$, and we simulate $x_1,c,R_v,$ (Section \ref{sec:Model}) as flat distributions; $E(B-V)$ is simulated with an exponential distribution following $\tau = 0.5$. 

\section{Model Overview}\label{sec:Model}

In this section, we review the light curve fitting that underlies the Stjörnumál code, and provide a basic overview of the dust model machinery. 

\subsection{SALT}

Lightcurves of SNe Ia are fit with an SED fitting model for use in cosmology analyses; here we focus on the SALT model as presented in \cite{Guy07,Guy10} and updated in \cite{Taylor21,Kenworthy21}. We use the SALT surface trained in \cite{Dovekie}.

The flux of an SN Ia is given by SALT as 
\begin{align} 
\label{saltmodel}
\begin{split}
F(\rm{SN}, p, \lambda) = x_{0} &\times\left[M_{0}(p, \lambda)+x_{1} M_{1}(p, \lambda)+\ldots\right] \\
&\times \exp [c C L(\lambda)],
\end{split}
\end{align}
where $x_0$ is the amplitude of the lightcurve, $x_1$ (`stretch') is the observed decay time of the light curve, and $c$ is the parameter encapsulating the colour of the SN Ia. In contrast, $M_0$, $M_1$, and $CL(\lambda)$ are global model parameters determined during the training: $M_0$ is the average 
Spectral Energy Distribution (SED) at each phase, $M_1$ is the variation within the SED, and $CL(\lambda)$ is the average colour law. 

Distances $\mu$ are then inferred via the Tripp estimator \citep{Tripp98} by
\begin{equation}
\label{eq:tripp}
    \mu = m_B + \alpha_{\rm SALT} x_1 - \beta_{\rm SALT} c - M_0
\end{equation}
where $m_B = -2.5\textrm{log}_{10}(x_0)$, $x_1$ and $c$ are previously defined, and the global nuisance parameters $\alpha_{\rm SALT2}$ and $\beta_{\rm SALT2}$ describe the stretch-luminosity and colour-luminosity relationships respectively. $M_0$ is the absolute magnitude of an SNe Ia with $x_1=c=0.$ 

In cosmology analyses, this distance modulus $\mu$ is corrected for additional, known sources of bias and selection. However, these bias corrections require a model of SN Ia scatter that relate SN Ia and host galaxy properties to the distance modulus.

In 2010, \cite{Kelly10,Sullivan10,Lampeitl10} discovered an anomalous post-standardisation difference in peak luminosity between SNe Ia in low-mass ($< {10} M_{*}$) and high-mass ($< {10} M_{*}$) galaxies, now dubbed `the mass step' $\gamma$. The $\gamma$ parameter is now commonly added as a post-hoc correction following a Heaviside function, but the astrophysical origin is still unknown. 

The nature of the mass step remains unsolved to this day, but explanations lay broadly in two camps: a byproduct of differing progenitor paths, or the dimming of the light curve by external dust. 

\subsection{Dust Model Overview}

Here we present a short review of the dust models employed in recent cosmological analyses. In short, these dust models bifurcate observed SNe Ia colour behaviour into two components: a component intrinsic to the SNe Ia $c_{\rm int}$, and an external dust component described by $R_V$ and $E(B-V)$.

The observed SALT colour of SNe Ia is then modeled as
\begin{equation}
    c_{\rm obs} = c_{\rm int} + E_{\rm Dust} + \epsilon_{\rm noise}
\label{eq:cobs}
\end{equation}
where $E_{\rm Dust}$ is the $E(B-V)$ reddening, and $\epsilon$ is additional and otherwise unaccounted for measurement noise.

The peak brightness of the SNe Ia is then modified by external dust as 
\begin{equation}\label{eq:deltamb}
    \Delta m_B = \beta_{\rm int} c_{\rm int} + (R_V + 1) * E_{\rm Dust} +\epsilon
\end{equation}

Please note that $\beta_{\rm int}$ and $\beta_{\rm SALT}$ are different parameters:  $\beta_{\rm SALT}$ is determined from a global fit of the SALT parameters and is a convolution of $\beta_{\rm int}$ and dust effects, whereas $\beta_{\rm int}$ is an intrinsic SN Ia feature.

Table \ref{tab:modelcomp} shows the assumed distributions of each parameter in the nominal dust model; while $c_{\rm int}$, $\beta_{\rm int}$, and $R_V$ are all assumed to be Gaussian, $E_{\rm dust}$ assumes the following distribution
\begin{equation}
    P(E_{\rm dust}) = 
    \begin{cases} 
      ~\tau_{E}^{-1}e^{-E_{\rm dust}/\tau_E} & ,~E_{\rm dust} > 0\\
      ~0 & ,~E_{\rm dust} \leq 0
   \end{cases}
\label{eq:dustprob}
\end{equation}
following \cite{Riess96} and \cite{Jha07}. 

In the case that the $R_V$ populations differ across host galaxy stellar mass, a colour-dependent magnitude difference may arise. The original \cite{BS20} proposed that $\overline{R}_V$ is different above and below $10M_*$, leading to a magnitude difference between low- and high-mass galaxies, thereby explaining the observed mass step as a dust phenomenon. 

\section{Methodology}\label{sec:Methodology}

The Stjörnumál pipeline works in two stages: simulation and training. With a selected model, a `simulation bank' is loaded in, and the desired number of simulations are generated from the priors via importance sampling (Section \ref{sec:Methodology:subsec:ImportanceSampling}) and saved. We employ no SBI at this step, but save the model parameters $\theta$ and the associated data to those model parameters, $d$. The data that we save is $\{ c,m_B,x_1,z,c_{\rm ERR}, m_{B\rm ERR},x_{1_{\rm ERR}}, \mu, M_{*}, \mu_{\rm RES} \}$ where $$\mu_{\rm RES} = \mu_{\rm model} - \mu_{\rm obs}$$
$\mu_{\rm model}$ is defined as in Section \ref{sec:Data:subsec:Sims}, $\mu_{\rm obs}$ is the SN Ia distance from Equation \ref{eq:tripp}, and $M_{*}$ is the log stellar mass of the host galaxy. We calculate $\mu_{\rm RES}$ once before the importance sampling; when doing so,  we assume $\beta = 3.1$ and $\alpha =0.145$, following the original DES5YR results. We cannot re-sample $\beta$ and $\alpha$ due to limitations in the current architecture. 

It is these distributions, $\{ c,m_B,x_1,z,c_{\rm ERR}, m_{B \rm ERR}, x_{1_{\rm ERR}}, \mu, M_{*}, \mu_{\rm RES} \}$, that Stjörnumál conditions on during the training step. The SALT parameter errors $c_{\rm ERR}, m_{B\rm ERR},x_{1_{\rm ERR}}$ are used by the network (Section \ref{sec:SBI}) to weigh signal from the SALT parameters; $z$ and $\mu$ are given to construct the cosmological relationship, and $\mu_{\rm RES}$ is given to `bootstrap' the relationship between model parameters and observed $\mu_{\rm RES}$ behaviour. The neural network is capable of learning the Tripp relation (\ref{eq:tripp}) from just the $x_1, c,m_B,\mu_{\rm theory}$ distributions, but this process is time consuming. Instead, we provide the relationship between $\mu$ and $\mu_{\rm res}$ explicitly, to expedite the inference process. We do not provide covariance between SALT parameters. The absolute magnitude of a supernova $M_0$ is fixed to be the same in both simulations and data. 

During training, the simulations are loaded, and a neural network (Section \ref{sec:SBI}) is fed $\{\theta,d\}$. Through various iterations, the network learns the relationships between the input model parameters $\theta$ and the output data $d$, which allows it to estimate the posterior distribution for a given set of real data. 

\subsection{Importance Sampling}\label{sec:Methodology:subsec:ImportanceSampling}

We keep much of the same methodology of the original \DTD, which we review here. \DTD, and hence Stjörnumál, use a `simulation bank' to importance sample from. This simulation bank is generated with \texttt{SNANA}, and includes the selection effects of the survey. The simulated $c, x_1, R_V$ distributions in the simulation bank are flat, with the exception of $\tau_E$, which is bound by the function $\tau^{-1}e^{E/\tau}$, with $\tau =0.5$ in this case. 

From the large simulation bank simulation (Section \ref{sec:Data:subsec:Sims}), we importance sample a selection of SNe Ia. For $\vec{\theta} = \theta_i$, each event is defined as 
\begin{equation}
    p = \prod^{n}_{i=1} p_i \\
\end{equation}
and
\begin{equation}
    p_{\rm ref} = \prod^{n}_{i=1} p_{\rm ref_i} \\
\end{equation}
where $p_i$ is the probability of the weighted simulation for parameter $i$, and $p_{\rm ref}$ is the same but for the input distribution. The weight function is then 
\begin{equation}
    P(X) = 
    \begin{cases}
       \textrm{if } \frac{p}{p_{\rm ref}} < U([0,1]), 0 \\
       \textrm{if } \frac{p}{p_{\rm ref}} \geq U([0,1]), 1 \\ 
    \end{cases}
\end{equation}
where $U([0,1])$ is a random number in a uniform distribution. The peak probability of each parameter is set to 1. At time of writing, Stjörnumál has native support for Gaussian, Double Gaussian, Logistic, and Exponential distributions, but more functions can be user-defined and added. The weights are then simultaneously applied to the simulation bank, which produces a simulacrum of the desired simulation with chosen inputs. This simulation is then down-weighted to match the size of the data for processing in SBI (see \ref{sec:SBI} for more).

Two parameters that do not follow the importance sampling framework within Stjörnumál: $\gamma$ and $\sigma_{\rm int}$. These parameters are not modelled as population-level parameters, and therefore need a special routine.

The mass step $\gamma$, which here refers only to the grey, achromatic magnitude step between low- and high-mass galaxies\footnote{Not the overall $\gamma$ as defined by \cite{Sullivan10}, but a specific component of it.}, is typically modelled as a Heaviside function or a quickly-changing sigmoid. Here we implement $\gamma$ as a Heaviside function, and after the importance sampling, $\gamma/2$ is added to $m_B ~\& ~\mu_{\rm RES}$ on either side of the user-defined split-location. 

The intrinsic scatter $\sigma_{\rm int}$ is additional grey scatter; we incorporate this as a Gaussian distribution of $\mathcal{N}(0, \sigma_{\rm int})$; after the importance sampling, this scatter is added to $m_B$ and $\mu_{\rm RES}$.

Stjörnumál has several error traps built-in: when $p/p_{\rm ref} > 1$, the resulting simulation is flagged as invalid and discarded from training; the same flagging occurs when there is not sufficient simulated SNe Ia to match the size of the data. More information on error traps and diagnostics is available in the github. 

\section{Models}\label{sec:Models}

Here we detail the models tested in this work. Our nominal model is shown in Table \ref{tab:modelcomp}, but in short resembles the original \cite{Popovic22} model with additional parameters for $\gamma$, $\sigma_{\rm int}$, and the $x_1$ distribution. Table \ref{tab:modelcomp} gives the overview of all of the models that we consider in our paper, and Table \ref{tab:priors} shows the associated priors. 

These models are broadly split into three categories - `Dust' models, including \textbf{Nominal}, \textbf{No Mass Step}, and \textbf{Logistic} $R_V$, the `Mixture Models' in \textbf{Banana Split} and \textbf{Banana Less Split}, and `Other' in \textbf{Single} $R_V$ and \textbf{2 Colour}.

The `Dust' models are variations on the \cite{BS20} model that posits varying $R_V$ distributions based on the host galaxy mass. In contrast, the `Mixture Models' do not explicitly tie results to the host galaxy stellar mass; instead, they model two independent populations of SNe that are well-tracked by the $x_1$ distribution.

\begin{table*}
    \centering
\begin{tabular}{c|lllccccc}
    Model & $c_{\rm int}$ & $R_V$ & $\beta_{\rm int}$ & $E(B-V)$ & $\gamma$ & $\sigma_{\rm int}$ & $x_1$ & $N_{\rm params}$ \\
    \hline
    \textbf{Nominal} & $\mathcal{N}(\mu,\sigma)$ & $\mathcal{N}(\mu,\sigma)$; ~~~ ${10} M_{*}$ & $\mathcal{N}(\mu,\sigma)$ & Exp$(\tau)$;  ${10} M_*$ & $\Theta$ & $\mathcal{N}(\sigma)$ & $\mathcal{N}(\mu,\sigma)$ & 14  \\
    \textbf{No Mass Step} & $\mathcal{N}(\mu,\sigma)$ & $\mathcal{N}(\mu,\sigma)$;  ~~~ ${10} M_{*}$ & $\mathcal{N}(\mu,\sigma)$ & Exp$(\tau)$; ${10} M_{*}$ & N/A & $\mathcal{N}(\sigma)$ & $\mathcal{N}(\mu,\sigma)$ & 13 \\
    \textbf{Logistic $R_V$} & $\mathcal{N}(\mu,\sigma)$ & $\mathcal{L}(L,k,\sigma)$ & $\mathcal{N}(\mu,\sigma)$ & Exp$(\tau)$; ${10} M_{*}$ & $\Theta$ & $\mathcal{N}(\sigma)$ & $\mathcal{N}(\mu,\sigma)$ & 13 \\
    \textbf{Single $R_V$} & $\mathcal{N}(\mu,\sigma)$ & $\mathcal{N}(\mu,\sigma)$ & $\mathcal{N}(\mu,\sigma)$ & Exp$(\tau)$ & $\Theta$ & $\mathcal{N}(\sigma)$ & $\mathcal{N}(\mu,\sigma)$ & 12 \\
    \textbf{2 Colour} & $\mathcal{N}(\mu,\sigma)$; ${10} M_{*}$ & $\mathcal{N}(\mu,\sigma)$ & $\mathcal{N}(\mu,\sigma)$;~~~ ${10} M_{*}$ & Exp$(\tau)$; ${10} M_{*}$ & N/A & $\mathcal{N}(\sigma)$ & $\mathcal{N}(\mu,\sigma)$ & 15 \\
    \textbf{Banana Split} & $\mathcal{N}(\mu,\sigma) \times 2$ & $\mathcal{N}(\mu,\sigma) \times 2$ & $\mathcal{N}(\mu,\sigma)$ & Exp$(\tau) \times 2$ & N/A & $\mathcal{N}(\sigma)$ & $\mathcal{N}(\mu,\sigma) \times 2$ &  $19$ \\
    \textbf{Banana Less Split} & $\mathcal{N}(\mu,\sigma) \times 2$ & $\mathcal{N}(\mu,\sigma)$ & $\mathcal{N}(\mu,\sigma)$ & Exp$(\tau)$ & N/A & $\mathcal{N}(\sigma)$ & $\mathcal{N}(\mu,\sigma) \times 2$ &  17 \\
\end{tabular}
    \caption{An overview of the models assessed in this paper, with the parameter distributions. Semi-colons denote a split on that parameter; the second entry indicates where this split is located. More information is available in the relevant subsections in Section \ref{sec:Models}. }
    \label{tab:modelcomp}
\end{table*}

\subsection{No Mass Step}

\begin{figure}
    \centering
    \includegraphics[width=8cm]{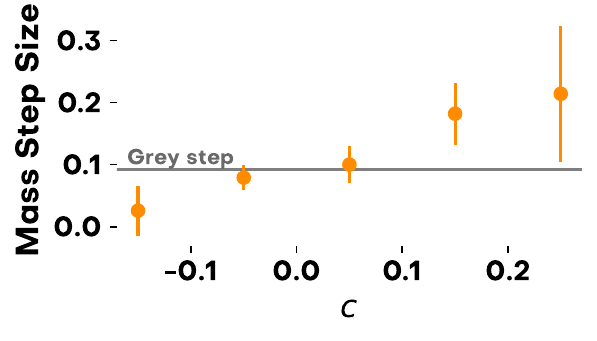}
    \caption{The DES-Dovekie mass step as a function of colour (orange points) and the `achromatic'/`grey' value .}
    \label{fig:MASSSTEP}
\end{figure}

The original \cite{BS20} and \cite{Popovic22} models ascribed the entirety of the observed mass step to $\Delta R_V$ across the host galaxy mass, and therefore did not allow for any achromatic difference in host galaxy magnitude. We show in Figure \ref{fig:MASSSTEP} an example of the colour-dependent step vs. an achromatic step for the DES5YR sample. While the nominal model includes an achromatic step $\gamma$, here we revert to this old definition and allow $\Delta R_V$ to account for the entirety of the mass step. 

\begin{table}
    \centering
    \begin{tabular}{c|ccc}
        Parameter & $\mu$ & $\sigma$ & $\tau$ \\
        \hline
        $c$ & $U$[$-0.1,0$] & $U$[$0.02,0.1$] & -- \\
        $R_V$ & $U$[$1.5, 6$] & $U$[$0.5,1.2$] & -- \\
        $E(B-V)$ & -- & -- & $U$[$0.05,0.5$] \\
        $\beta_{\rm int}$ & $U$[$1.5,3$] & $U$[$0.05,0.3$] & --   \\
        $x_1$ & $U$[$-2,2$] & $U$[$0.2, 1.5$] & -- \\
        \hline
        Parameter &  \multicolumn{3}{c}{Range}  \\
        $\gamma$ & \multicolumn{3}{c}{$U$[$-0.2,0.2$]} \\
        $\sigma_{\rm int}$ & \multicolumn{3}{c}{$U$[$0,0.1$]}   \\ 
    \end{tabular}
    \caption{Priors for our hyper parameters. }
    \label{tab:priors}
\end{table}

\subsection{Logistic $R_V$}

\cite{Popovic24b} proposed a logistically-distributed $R_V$ population modelled as:
\begin{equation}\label{eq:logistic}
    R_V = \frac{L}{1+e^{k(\log(M_*) - 10)}}
 +2\end{equation}
with two parameters $k,L$. They found promising results, but did not retrain a model with this new distribution. We do so here, fitting this logistic function with an additional scatter $\sigma$ around the central sigmoid, reducing the number of parameters from two distinct Gaussians (4) to only 3 parameters: $L,k,\sigma$.

\begin{figure}
    \centering
    \includegraphics[width=8cm]{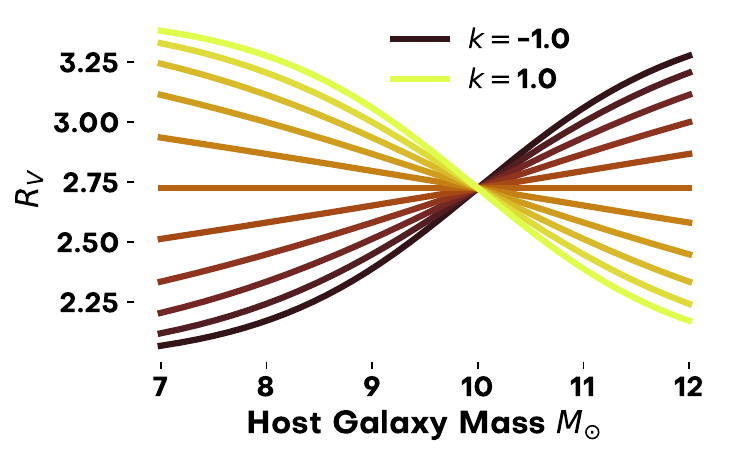}
    \caption{An example of varying $k$ within the logistic function. When $k=1$, the logistic function reduces to a sigmoid. }
    \label{fig:LOGISTICEXAMPLE}
\end{figure}

Figure \ref{fig:LOGISTICEXAMPLE} shows an example of varying the $k$ parameter for a fixed $L$, demonstrating the smoothly-varying nature of this function compared to the historical 2-Gaussian approach. 

\subsection{Single $R_V$}

\cite{BS20} also provided an initial exploration of the impact of $R_V$ on SN Ia but without a split on host-galaxy mass. 
For this model, we take the nominal model but do not split $R_V$ and $E(B-V)$ on the host logmass. Instead, we fit a single dust distribution for the entire sample, but allow the achromatic mass step.

\subsection{Two $c_{\rm int}$ Distributions}
Rather than a set of host galaxy dependent $R_V$ values, \cite{Gonzalez-Gaitan21} suggest that the observed $\beta_{\rm SALT}$ mass dependence, and subsequent mass step, arise from two distinct populations of SNe with different $c_{\rm int}$ and $\beta_{\rm int}$, separated by host galaxy stellar mass, effectively
\begin{equation}
    \Delta m_B  = 
    \begin{cases}
       \textrm{if } M_{*} < 10, \beta_{\rm SN,1} c_{\rm int,1} + (R_V + 1) * E_{\rm Dust} +\epsilon \\
       \textrm{if } M_{*} \geq 10, \beta_{\rm SN,2} c_{\rm int,2} + (R_V + 1) * E_{\rm Dust} +\epsilon \\ 
    \end{cases}
\end{equation}
where $\beta_{\rm SN,1} c_{\rm int,1}$ refer to the low mass intrinsic population, and $\beta_{\rm SN,2} c_{\rm int,2}$ is the high mass population. 

To implement this model, we take the Single $R_V$ model and then split the $\beta_{\rm int}$ and $c_{\rm int}$ populations on a host galaxy stellar mass of $10 M_{*}$.

\subsection{{Banana Split}}

Inspired by \cite{Wojtak23}, \cite{Rubin26} create the `Banana Split' model, allowing for two distinct populations of SNe Ia that are, as in \cite{Wojtak23}, generally tracked by differences in the SN Ia stretch distribution. The Banana Split model is given as
\begin{align}
m^{\rm model, F/S}_B
    &= -\alpha^{F/S}(x^{\rm True}_1 - x^{F/S}_1)
    + \beta_B c^{\rm True}_B \nonumber \\
    &\quad + [\beta^{\rm low}_R(1-P) + \beta^{\rm high}_R P]c^{\rm True}_R
    - \gamma + \mu_{\rm cosmo}
\label{eq:bananasplit}
\end{align}
where the SNe Ia are split into two populations, fast or slow ($F$ or $S$). In either case, both $x_1$ modes are modelled as a Gaussian, and the colour modelling of Banana Split is similar to the $c_{\rm int}+E_{\rm Dust}$ modelling detailed earlier in this paper; essentially, two copies of the single $R_V$ model. 

Unlike the original paper, we do not model two independent $\alpha$ for each population - this is not possible with the current architecture. Instead, we use the same $\alpha$ as before. 

Here we enter the mixture models. Previous sections have tied SN populations explicitly to host galaxy parameters; the mixture models do not. To implement this model, we first ensure that there is no degeneracy between the two models. This is done by changing the $x_1$ prior: Population A has $\overline{x}_1 \in [-2,0)$, and Population B has $\overline{x}_1 \in (0,2]$, thereby preventing extreme degeneracies when the populations `cross over'. Similarly, we set $\beta_{\rm int}$ as a shared parameter across the two populations; both of these decisions follow \cite{Rubin26}. The mixture model mode of Stjörnumál includes a $f_{\rm mix}$ parameter that details the relative weight of the two populations: a value of 1 or 0 indicates a single population. Additionally, we change the definition of $\gamma$: rather than a magnitude offset between the low- and high-mass host galaxies, when the mixture mode is enabled, $\gamma$ becomes a magnitude offset between the two populations, mimicking a difference in absolute brightness $M_0$ values.

\subsection{{Banana Less Split}}

The Banana Split model still allows for differing $R_V$ and 
$E(B-V)$ distributions across the two populations. To isolate the potential impacts of solely differing SN populations, we modify the Banana Split model and set $R_V$ and $E(B-V)$, alongside $\beta_{\rm SN}$, as a single population, dubbing this new model `Banana Less Split'\footnote{As it is, indeed, less split than Banana Split.} This alternative reduces the total parameter space for external properties by half.


\section{SBI}\label{sec:SBI}
{Bayesian inference involves updating prior belief about the parameters of interest through adopting a likelihood and then using Bayes' rule. For parameters $\boldsymbol{\theta}$ and data $\boldsymbol{d}$, Bayes' rule states
\begin{equation}
p(\boldsymbol{\theta} \mid \boldsymbol{d}) 
= \frac{p(\boldsymbol{d} \mid \boldsymbol{\theta}) \, p(\boldsymbol{\theta})}{p(\boldsymbol{d})},
\end{equation}
where $p(\boldsymbol{d} \mid \boldsymbol{\theta})$ is the likelihood, 
$p(\boldsymbol{\theta})$ is the prior distribution, and 
$p(\boldsymbol{d})$ is the evidence: $
p(\boldsymbol{d}) 
= \int p(\boldsymbol{d} \mid \boldsymbol{\theta}) \, p(\boldsymbol{\theta}) \, d\boldsymbol{\theta}.
$ }

{Simulation based inference (SBI) provides a solution to Bayesian inference for stochastic models $p(\boldsymbol{x}\mid \boldsymbol{\theta})$  when the likelihood is intractable but can be sampled from using a simulator model. In this paper we focus on methods that target the posterior and the likelihood-to-evidence ratio $p(\boldsymbol{d} \mid \boldsymbol{\theta})/p(\boldsymbol{d})$ known as neural posterior estimation (NPE) and neural ratio estimation (NRE), respectively.}

We employ NPE and NRE for obtaining our posteriors and comparing our models, respectively. Specifically, we use the \texttt{sbi} package in python \citep{SBIPackage}. To train the model, we condition on $\{ c,m_B,x_1,z,c_{\rm ERR}, x_{1_{\rm ERR}}, \mu, M_{*}, \mu_{\rm RES} \}$. For each model, we simulate 100,000 simulations; each contains the same number of data points as the data that will be conditioned on. Therefore, each model is tailored specifically to a single dataset and posterior models must be retrained when investigating new data. This specificity is somewhat ameliorated by quick simulation and training times; simulating 100,000 data sets takes $\sim5$ minutes, and training a model on these 100,000 data sets takes $\sim4$ hours. Model training is done in chunks to save memory, and the loss is reported at each step of 10,000 batches. 

\subsection{Neural Posterior Estimation}

NPE involves specifying a conditional density estimator 
$q_{\boldsymbol{\phi}}(\boldsymbol{\theta} \mid \boldsymbol{d})$ together with a neural summary statistic (NSE) 
$S_{\boldsymbol{\omega}}(\boldsymbol{d})$ parametrised by 
$\boldsymbol{\phi}$ and $\boldsymbol{\omega}$, respectively. The parameters 
$\boldsymbol{\phi}$ and $\boldsymbol{\omega}$ of these models are learned by minimising the expected forward 
Kullback--Leibler (KL) divergence between the true posterior and its approximation, which is equivalent to 
\begin{equation}
\mathcal{L}_{\mathrm{NPE}}(\boldsymbol{\phi}, \boldsymbol{\omega})
= - \mathbb{E}_{\boldsymbol{d}, \boldsymbol{\theta}}
\big[
\log q_{\boldsymbol{\phi}}(\boldsymbol{\theta} \mid S_{\boldsymbol{\omega}}(\boldsymbol{d}))
\big],
\end{equation}
up to a term that does not depend on $\boldsymbol{\phi}$ and $\boldsymbol{\omega}$ and the expectation is taken over samples from the simulator. 

\subsection{Neural Ratio Estimation}

NPE does not provide any way to compare models. Standard neural ratio estimation (NRE) uses neural networks to approximate the likelihood-to-evidence ratio. In this paper, we opt instead to target the likelihood ratio between two different models and approximate
\begin{equation}\label{eq:ratio}
r_{\boldsymbol{\phi}}(\boldsymbol{\theta}, S_{\boldsymbol{\omega}}(\boldsymbol{d})) 
\approx p(\boldsymbol{d} \mid M_1)/p(\boldsymbol{d} \mid M_2).
\end{equation}
where $M_1$ and $M_2$ refer to two different models and $S_{\boldsymbol{\omega}}$ refers to our neural embedding statistic similarly to in NPE. The Bayes ratio $p(\boldsymbol{d} \mid M_1)/p(\boldsymbol{d} \mid M_2)$ allows us to perform model comparison.

To estimate this ratio, we formulate density-ratio estimation as a binary
classification problem that distinguishes samples drawn from
$p_1(\boldsymbol{d},\boldsymbol{\theta})
=
p(\boldsymbol{\theta}\mid M_1)\,
p(\boldsymbol{d}\mid \boldsymbol{\theta}, M_1)$
and
$p_2(\boldsymbol{d},\boldsymbol{\theta})
=
p(\boldsymbol{\theta}\mid M_2)\,
p(\boldsymbol{d}\mid \boldsymbol{\theta}, M_2)$,
yielding the objective
\begin{equation}
\begin{aligned}
\mathcal{L}_{\mathrm{NRE}}(\boldsymbol{\phi}, \boldsymbol{\omega})
&=
- \mathbb{E}_{p_1(\boldsymbol{d}, \boldsymbol{\theta})}
\left[
\log \sigma\!\left(
f_{\boldsymbol{\phi}}(\boldsymbol{\theta},
S_{\boldsymbol{\omega}}(\boldsymbol{d}))
\right)
\right]\\
& \quad\,\,
-
\mathbb{E}_{p_2(\boldsymbol{d}, \boldsymbol{\theta})}
\left[
\log \left(
1-\sigma\!\left(
f_{\boldsymbol{\phi}}(\boldsymbol{\theta},
S_{\boldsymbol{\omega}}(\boldsymbol{d}))
\right)
\right)
\right]
].
\end{aligned}
\end{equation}

We integrate the NRE functionality using the same simulator software, but compare the efficacy of our models using a binary classifier to calculate the Bayes Factor (BF, Eq. \ref{eq:ratio}). We estimate the true temperature of the classifier after classification following \cite{Guo2017}; we take the output of the binary classification $z_{\rm CLAS}$ and rescale by $T$ to better quantify our errors. For each model, we calculate the Bayes Factor 5 times, and take the median value, following \cite{Trotta08} to interpret the significance.

\subsection{Neural Summary Statistic}

To construct a permutation-invariant summary statistic for i.i.d.\ observations, we adopt a Deep Sets architecture \citep{2017arXiv170306114Z}. Given a set $\boldsymbol{d} = \{d_i\}_{i=1}^N$, our Deep Set architecture takes the form 
\begin{equation}
S_{\boldsymbol{\omega}}(\boldsymbol{d}) 
= \rho_{\boldsymbol{\omega}}\!\left(
\sum_{i=1}^N 
\frac{\exp\!\big(\phi_{\boldsymbol{\omega}}(d_i)\big)}
{\sum_{j=1}^N \exp\!\big(\phi_{\boldsymbol{\omega}}(d_j)\big)}
\, \phi_{\boldsymbol{\omega}}(d_i)
\right).
\end{equation}
Where $\rho$ and $\phi$ are feedforward neural networks.

Contrary to the original Dust2Dust, then, we do not mock an explicit likelihood function (see Equations 20,21 in \cite{Popovic22}). Rather, our neural network approximates the posterior by minimising the loss function with use of the embedding network we employ. 

\section{Validation}\label{sec:Validation}

Before providing results, we test the accuracy of Stjörnumál through three tests of the posterior, comprising Section \ref{sec:Validation:subsec:Validation}: Tests of Accuracy with Random Points (TARP, Section \ref{sec:Validation:subsubsec:TARP}, Simulation Based Calibration (SBC, Section \ref{sec:Validation:subsubsec:SBC}), and Predictive Posterior Checks (PPC, Section \ref{sec:Validation:subsubsec:PPC}). In Section \ref{sec:Validation:subsec:SNANA}, we compare the predictions and performance of Stjörnumál to simulations generated with \texttt{SNANA}. We test the cosmological dependency of our model in Section \ref{sec:Validation:subsec:Cosmo}. Finally, we test internal model mis-specification in Section \ref{sec:Validation:subsec:Miss}.

\subsection{Posterior Validation}\label{sec:Validation:subsec:Validation}

This subsection focuses on evaluating the internal efficacy and accuracy of the Stjörnumál pipeline. Broadly, subsections \ref{sec:Validation:subsubsec:TARP}, \ref{sec:Validation:subsubsec:SBC}, and \ref{sec:Validation:subsubsec:PPC} all focus on different methods of generating simulated data and estimating our ability to recover the input values within error, inside the Stjörnumál pipeline. 

\begin{figure}
    \centering
    \includegraphics[width=8cm]{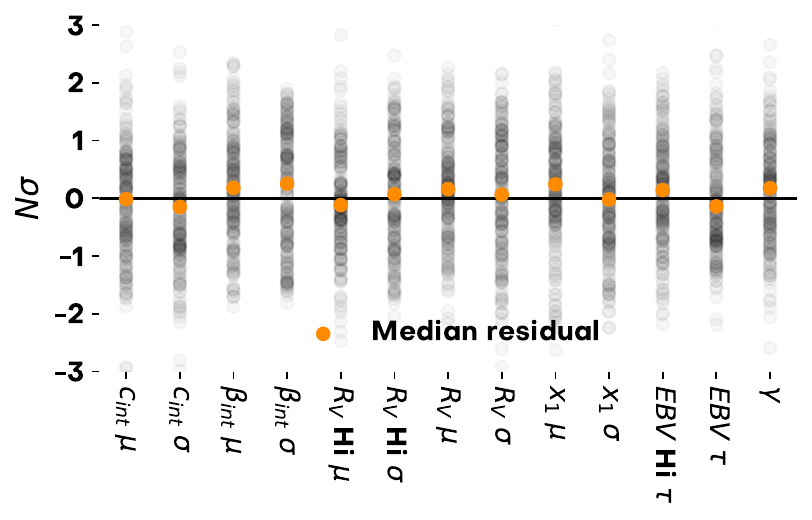}
    \caption{Frequentist calibration done for Stjörnumál over 200 simulations, showing how well the input parameters are inferred. Black points show individual realisations, and orange points are the mean of the 200 simulations. Orange rectangle shows the $1-\sigma$ model uncertainty. A black line is plotted to show $N \sigma = 0$.}
    \label{fig:Frequentist_Stjarna}
\end{figure}

Before moving into the coverage plots, Figure \ref{fig:Frequentist_Stjarna} shows the frequentist $N\sigma$ results from running Stjörnumál on 200 samples generated from the simulation bank. We see excellent internal consistency in the model. 

\subsubsection{Tests of Accuracy with Random Points}\label{sec:Validation:subsubsec:TARP}

Tests of Accuracy with Random Points (TARP) by \cite{TARP} provides a quick method of estimating posterior coverage with SBC. Given a random test set of parameters and data $(\theta,x)$ and a set of reference parameters $\theta_r$, we draw posterior samples $d$ given each $\theta$, calculate the distance $r$ between each $\theta$ and $\theta_r$, and count how many posterior samples have a distance to $\theta_r < r$. A more in-depth explanation is given in \cite{TARP}; for us, it suffices to say that a perfect correlation between the credibility level $\alpha$ and the Expected Coverage Probability corresponds to a well-calibrated and trustworthy posterior. 

\begin{figure}
    \centering
    \includegraphics[width=8cm]{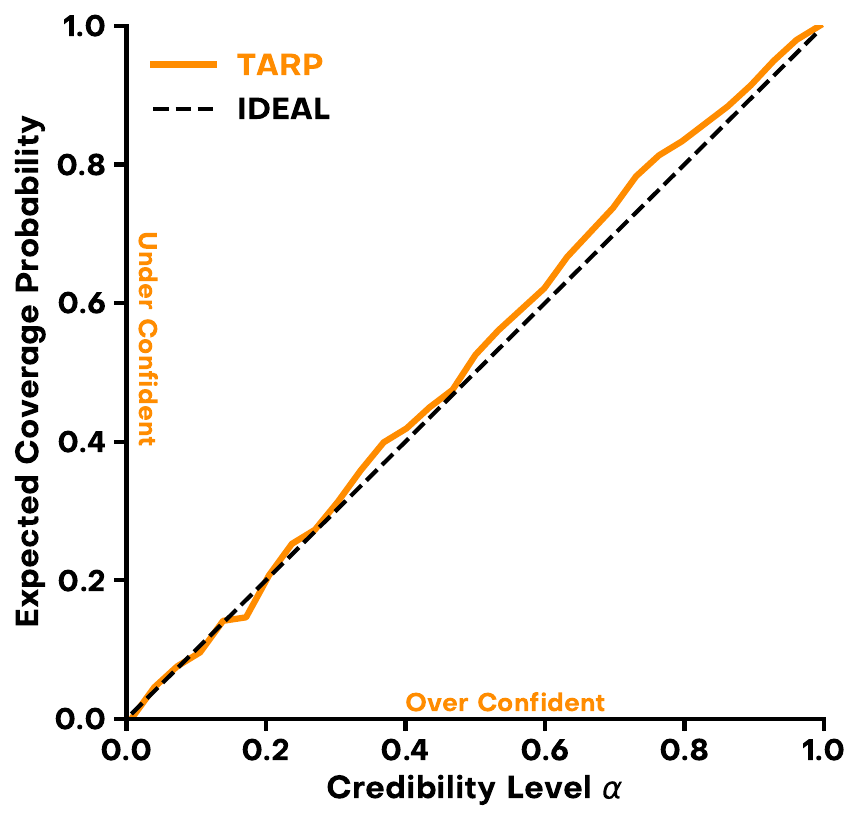}
    \caption{Stjörnumál frequentist calibration of posteriors after 4,000 draws from 200 repeated experiments. The black dashed line is the exact posterior; the orange solid line is the result from Stjörnumál. The area under the exact posterior line indicates predicted posteriors that are too narrow; the region above the exact posterior line indicates a posterior that is too broad. The TARP calibration shows an excellent estimation of the posterior. }
    \label{fig:TARP}
\end{figure}

Figure \ref{fig:TARP} shows our TARP results, from 4000 posterior samples across 200 realisations. Regions under the 1:1 line indicate an over confident posterior, those above indicate an under confident one. Less succinctly said, the TARP being in the region above the 1:1 line demonstrates that errors are overestimated and likely inflated, and below the 1:1 line, the errors are underestimated, making the SBI output overly confident in its posterior estimation. We see that our nominal model performs excellently. 

\subsubsection{Simulation Based Calibration}\label{sec:Validation:subsubsec:SBC}


TARP, while visually intuitive, relies on additional hyper parameters to describe the behaviour of the posterior. We also employ an additional method that does not require any additional parameters: Simulation Based Calibration \citep{SBC}. With some ground truth $\tilde{\theta}$ and corresponding data $\tilde{x}$, we can integrate over the exact posteriors of our model:
\begin{equation}\label{eq:priorprior}
    \pi(\theta) = \int d\tilde{x}d\tilde{\theta}\pi(\theta|\tilde{x})\pi(\tilde{x}|\theta)\pi(\tilde{\theta}).
\end{equation}
The average of any exact posterior estimation with respect to the data generated from the joint distribution reduces to the corresponding prior expectation. Because of this, the prior sample $\tilde{\theta}$ and any posterior sample $\{ \theta_1,...,\theta_L\}$ will be distributed the same; for any 1D variable $\mathcal{f} \Theta \xrightarrow{} \mathbb{R}$, the rank statistic of the prior relative to the posterior
\begin{equation}\label{eq:SBC}
    r(\{ \mathcal{f}(\theta_1),...,\mathcal{f}(\theta_L) \}, f(\tilde{\theta})) = \sum^L_{l=1}\mathbb{I}[\mathcal{f}(\theta_l) < f(\tilde{\theta})] \in [0,L]
\end{equation}
will be uniformly distributed.


This leads well into calibration of the posteriors: a `bell shaped' distribution indicates that the computed posterior is \textit{wider} than the true posterior. Inversely, a `U shaped' distribution shows that the computed posterior is likely narrower than the true posterior. In the case of posteriors that are skewed to one side of the distribution, the posterior for that given parameter is likely outside the prior range. 

\begin{figure}
    \centering
    \includegraphics[width=9cm]{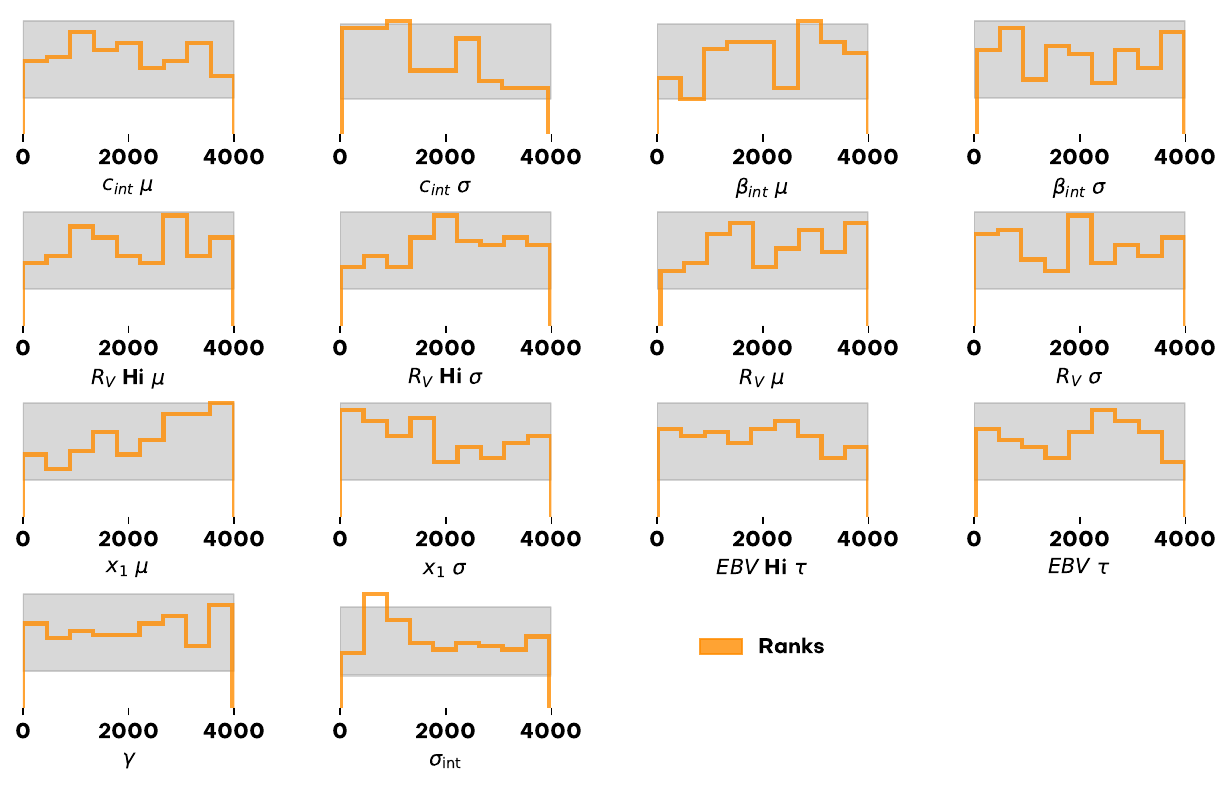}
    \caption{Rank histograms (orange) for the Stjörnumál fiducial model. The coverage expected under uniformity is shown in grey fill; a flat histogram within the bounds of this grey fill indicates well-calibrated posteriors. Bell or U-shaped distributions indicate a posterior that is too wide or narrow, respectively. Our posteriors are generally well-calibrated.}
    \label{fig:rank}
\end{figure}

To perform SBC, we make use of the built-in SBC package from SBI, \texttt{run\_sbc}. We sample 200 $\tilde{\theta}$ draws from our prior, and for each of these draws, simulate 200 $\tilde{x}$ each corresponding to their respective $\tilde{\theta}$. For each $\tilde{x}$ we use our Stjörnumál posterior to draw 4,000 samples; the corresponding rank statistics (Equation \ref{eq:SBC}) are then binned and plotted in Figure \ref{fig:rank}.

\begin{figure}
    \centering
    \includegraphics[width=8cm]{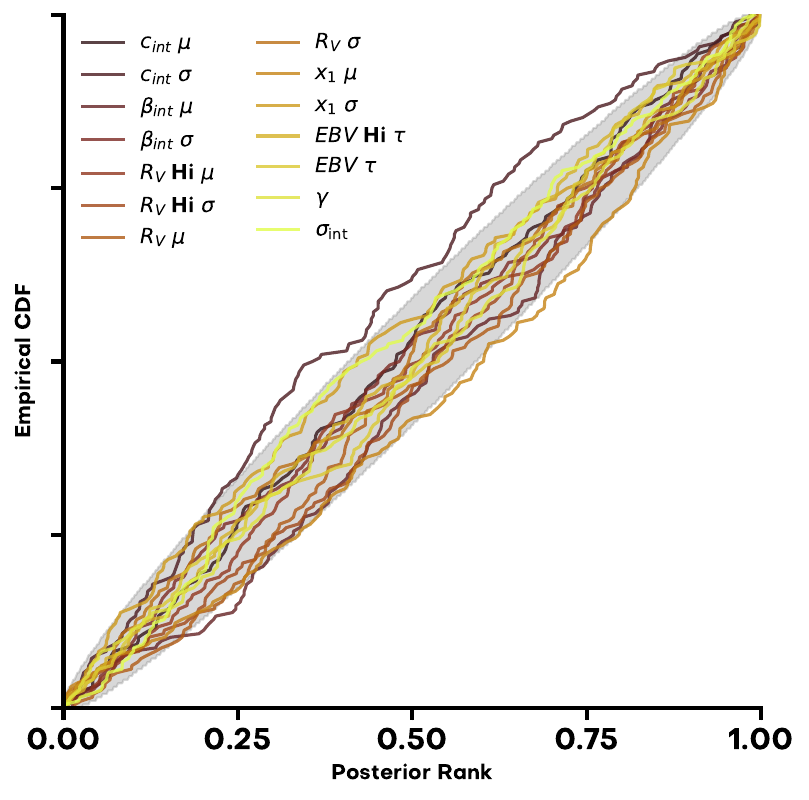}
    \caption{Stjörnumál frequentist calibration for each model parameter, repeated 200 times with 4,000 draws from the posterior. The grey shaded area is the 95\% confidence band under perfect calibration. We see excellent posterior calibration. }
    \label{fig:rankcdf}
\end{figure}

Alternatively, the SBC rank plots can be converted into a continuous distribution function; this visualisation allows us to compare the rank plots directly to the TARP, but for individual parameters. We show this characterisation in Figure \ref{fig:rankcdf}.

We see good posterior calibration for our parameters with the exception of  $\sigma x_1$, which lays slightly out of the expected range. Nonetheless, this does not impact our TARP nor PPC.

\subsubsection{Posterior Predictive Checks}\label{sec:Validation:subsubsec:PPC}

We assess the Stjörnumál model in replicating the observed data via our posterior predictive checks. We sample from the posterior 5,000 times, and simulate 5,000 datasets from these posterior draws. We compare only the distributions that Stjörnumál is able to affect: $\{c,m_B,x_1,\mu_{\rm RES} \}$, and compute a Wasserstein distance between the two distributions to check the goodness of fit. 

\begin{figure}
    \centering
    \includegraphics[width=6cm]{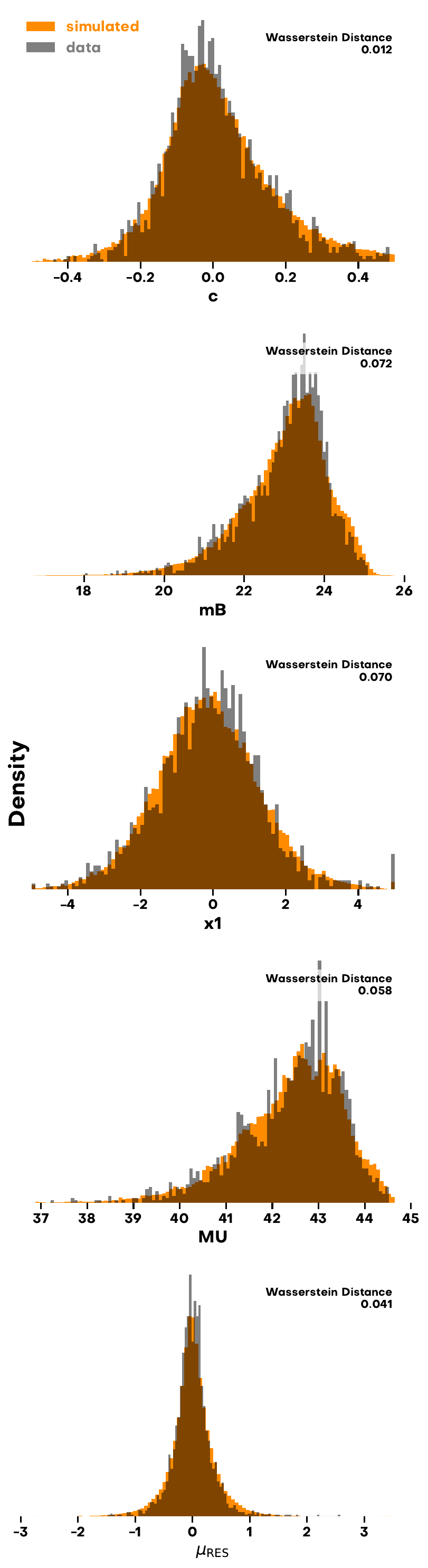}
    \caption{Stjörnumál simulated outputs (orange) and data (black) for ${c, m_B, x_1, \mu, \mu_{\rm RES}}$. A Wasserstein distance is given for each histogram; we see no Wasserstein distance $> 0.1$. }
    \label{fig:PPC}
\end{figure}

Figure \ref{fig:PPC} shows the results; we find Wasserstein distances consistently under 0.1, and see good agreement between the data and our posterior predictions. 

\subsection{Model Recovery on SNANA Simulations}\label{sec:Validation:subsec:SNANA}

Of course, our previous tests are only as valid as the Stjörnumál simulation framework, which is not used in cosmology analyses. We calculate the recovery of our input parameters by sampling from our priors 200 times, simulating with those parameters in \texttt{SNANA}, and checking how well Stjörnumál recovers the parameters. Due to \texttt{SNANA} architecture, we do not simulate $\gamma$ or $\sigma_{\rm int}$\footnote{For $\gamma$, \texttt{SNANA} requires either a bespoke WGTMAP or HOSTLIB; iterating over these requires 200 individual, bespoke simulation blocks.}.

\begin{figure}
    \centering
    \includegraphics[width=8cm]{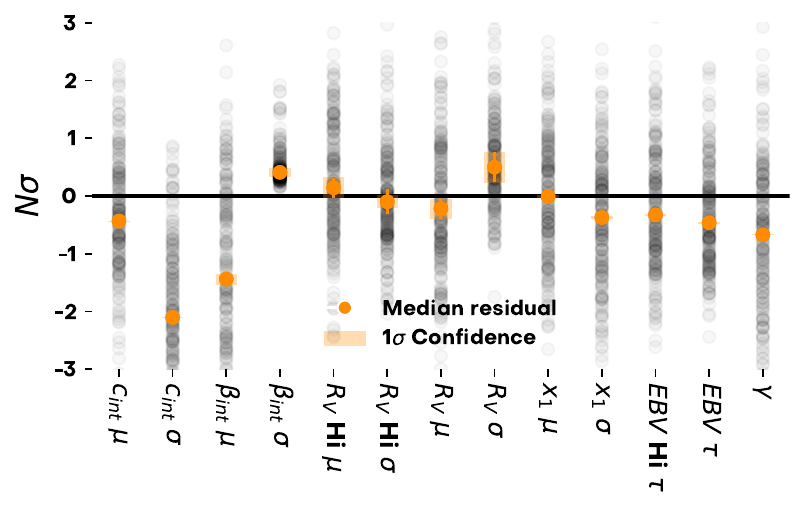}
    \caption{Frequentist calibration done for Stjörnumál over 200 \texttt{SNANA} simulations, showing how well the input parameters are inferred. Black points show individual realisations, and orange points are the mean of the 200 simulations. Orange rectangle shows the $1-\sigma$ model uncertainty. A black line is plotted to show $N \sigma = 0$.}
    \label{fig:Frequentist}
\end{figure}

Figure \ref{fig:Frequentist} shows the results of running Stjörnumál on the 200 \texttt{SNANA} simulations; we see one parameter with a bias $>2\sigma$: $c_{\rm int}~\sigma$. Interestingly, we find that $c_{\rm int} \sigma$ in \texttt{SNANA} simulations is consistently overestimated by $0.008$, independent of other parameters. 
We find a slight bias ($\sim1.5\sigma$) in $\beta_{\rm int }~\mu$, which is driven by the lower ends of the $\beta_{\rm int}$ range. Overall, performance on \texttt{SNANA} simulations is degraded compared to internal Stjörnumál simulations; this is not unexpected, but warrants further work. 

\subsection{Cosmological Dependence}\label{sec:Validation:subsec:Cosmo}

We have assumed $\Lambda$CDM cosmology in our simulator; here we test the validity of this assumption. Using our nominal Stjörnumál posterior, we perform the validation tests from \ref{sec:Validation:subsec:Validation} while varying the cosmology of our simulator, mocking the effects of cosmology mis-match between the data and the assumed simulated cosmology. We perform this test in steps of $w = 0.05$ from $w = -0.9 \xrightarrow{} -1.1$. 

\begin{figure}
    \centering
    \includegraphics[width=8cm]{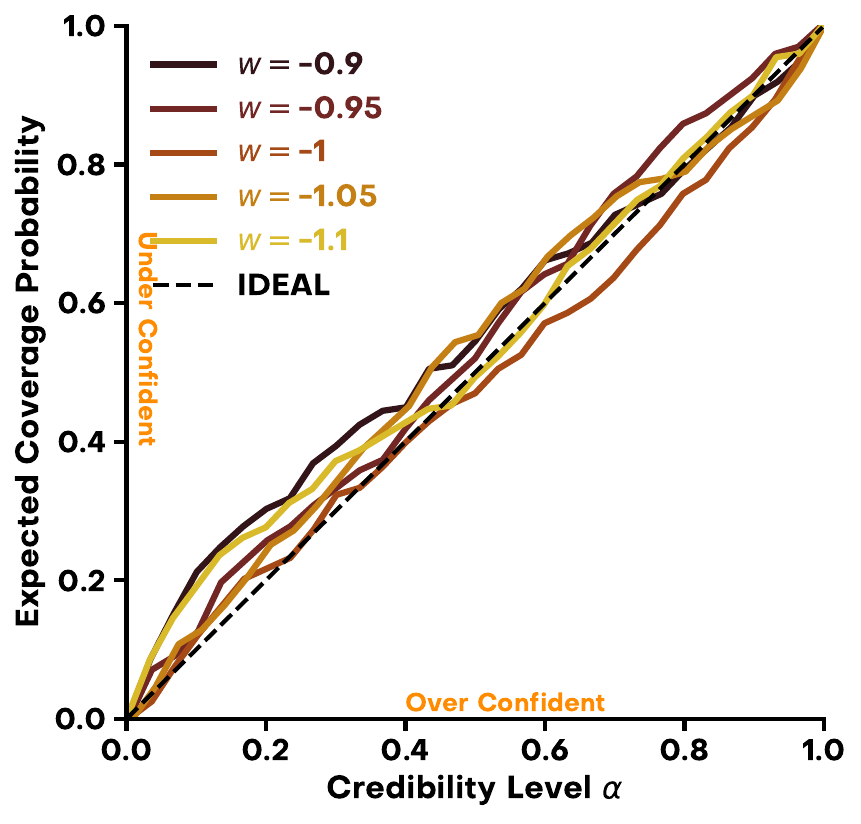}
    \caption{TARP plot for the Stjörnumál posteriors as a function of changing $w$ values. We see a slight dependence of our posterior confidence on the assumed cosmology. }
    \label{fig:tarpw}
\end{figure}

\begin{figure}
    \centering
    \includegraphics[width=9cm]{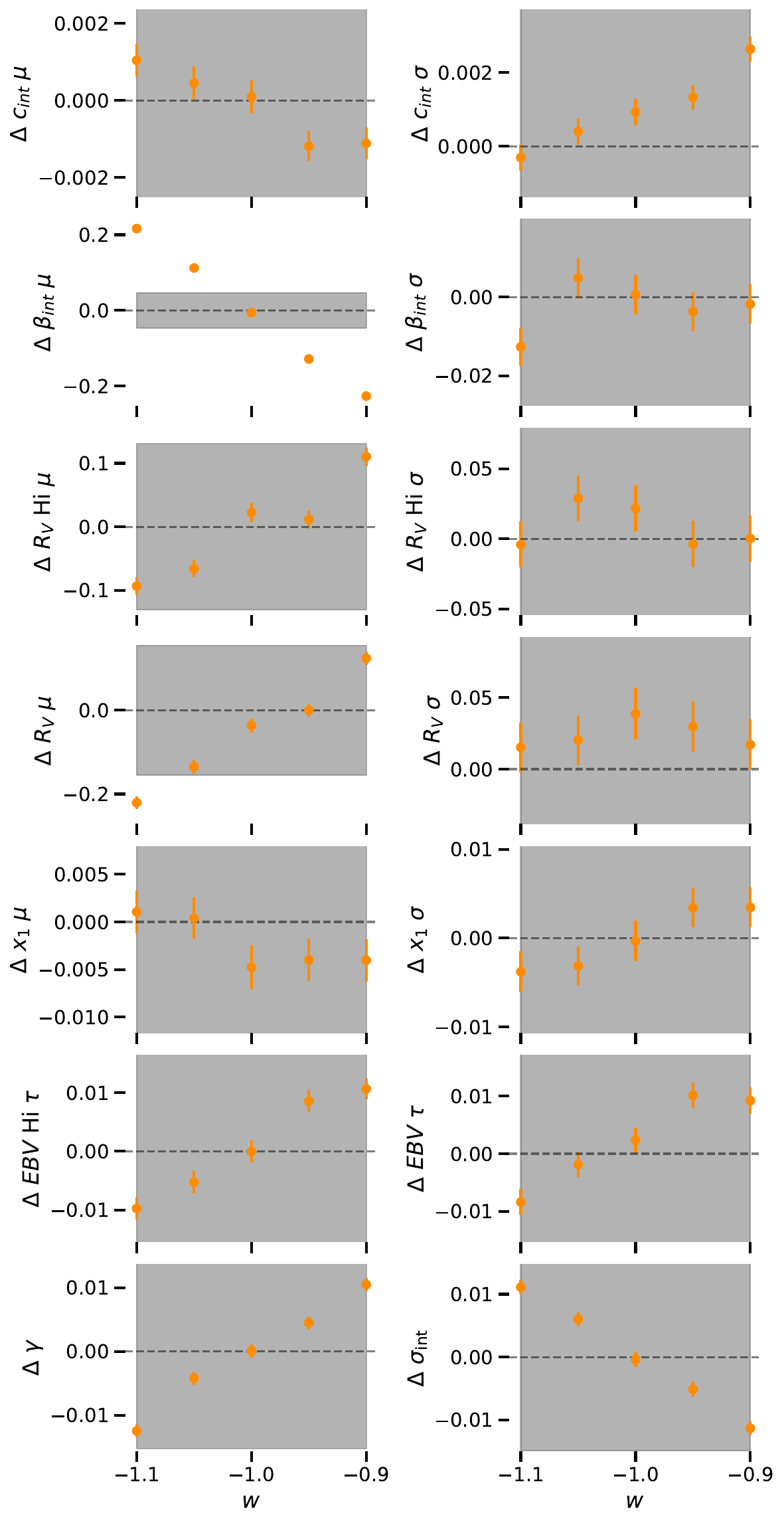}
    \caption{Bias in inferred parameters as a function of $\Delta w$ between the model and the simulated cosmology. Orange points show the average bias in the relevant parameter; errors include the $1/\sqrt{N}$ term. Grey fill indicates the $1\sigma$ uncertainty on that parameter as shown in Table \ref{tab:results}. }
    \label{fig:cosmoparams}
\end{figure}

\begin{figure*}
    \centering
    \includegraphics[width=18cm]{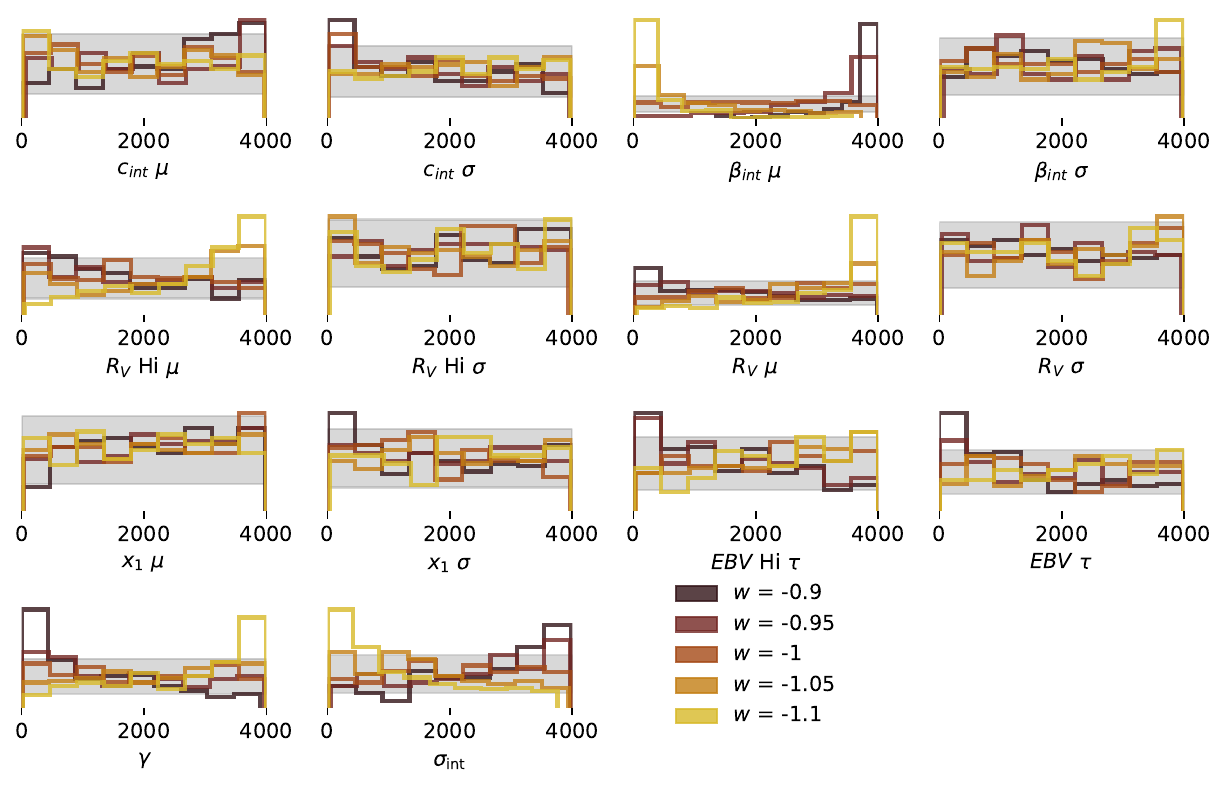}
    \caption{Rank histograms as a function of assumed cosmology for Stjörnumál. We colour-code the different assumed cosmologies, and see good model agreement for most of our parameters. }
    \label{fig:rankcosmo}
\end{figure*}

We see cosmological dependence of the model performance in Figure \ref{fig:tarpw}, especially at the extremes of our tested cosmology range. Figure \ref{fig:rankcosmo} shows the breakdown per-variable: we see strong cosmology dependence in $\beta_{\rm int}$ and the mass step $\gamma$, and some more minor dependence in the mean $R_V$ values. 

We complement this rank plot by showing the per-parameter change as a function of input cosmology in Figure \ref{fig:cosmoparams}. We draw 200 samples from the prior for each input cosmology, and calculate the average offset from the true value along with its associated error, which is then divided by $\sqrt{200}$. We find that our systematic shifts are less than the 1$\sigma$ uncertainties for the parameters, with the exception of $\beta_{\rm int} \mu$, which varies strongly with the cosmology. 

We have fixed $\beta_{\rm SALT} = 3.1$ when calculating $\mu_{\rm RES}$; we allow it to vary here. We find similar results to varying the input cosmology;
$\beta_{\rm int}~\mu$ and $R_V~\mu$ correlate strongly with the assumed $\beta_{\rm int}$, and are inferred outside the 1$\sigma$ value at $\Delta\beta_{\rm SALT} = 0.2$. We perform a similar test for $\alpha_{\rm SALT}$; both are shown in Appendix \ref{sec:appendix:sec:nuisance}.

\subsection{Model Misspecification}\label{sec:Validation:subsec:Miss}

In this section, we test the ability of Stjörnumál to discern between different models in its simulation pipeline. We do this with the NRE pipeline, and in each case, generate a $1\times$ model of DES within Stjörnumál of the desired model. This simulated data is then run through the NRE pipeline to compare the model it was generated with, and the nominal model. As for the data, we run the NRE pipeline 5 times, to minimise shot noise. 

\begin{table}
    \centering
    \begin{tabular}{c|c}
        Model & $\ln(\mathcal{Z})$ \\
        \hline
        \textbf{No Mass Step} & $-1.06 \pm 0.01$\\
        \textbf{Logistic $R_V$} & $-0.58 \pm 0.15$\\
        \textbf{Single $R_V$} & $-0.59 \pm 0.08$ \\
        \textbf{2 Colour} & $-0.95 \pm 0.32$ \\
        \textbf{Banana Split} & $-3.03 \pm 0.25$ \\
        \textbf{Banana Less Split} & $-2.99 \pm 0.43$ \\
    \end{tabular}
    \caption{Bayes Factors for models simulated in Stjörnumál and compared to the nominal model. A negative value indicates preference for the correct model. }
    \label{tab:Model_Miss}
\end{table}

Table \ref{tab:Model_Miss} shows the results. Stjörnumál is moderately well-able to differentiate between models, landing on $ \ln(\mathcal{Z})\approx 1$ across all the models. It performs best with identifying the two population models, and worse, surprisingly, with Single $R_V$ and Logistic $R_V$.

\section{Results}\label{sec:Results}

\begin{figure*}
    \centering
    \includegraphics[width=17cm]{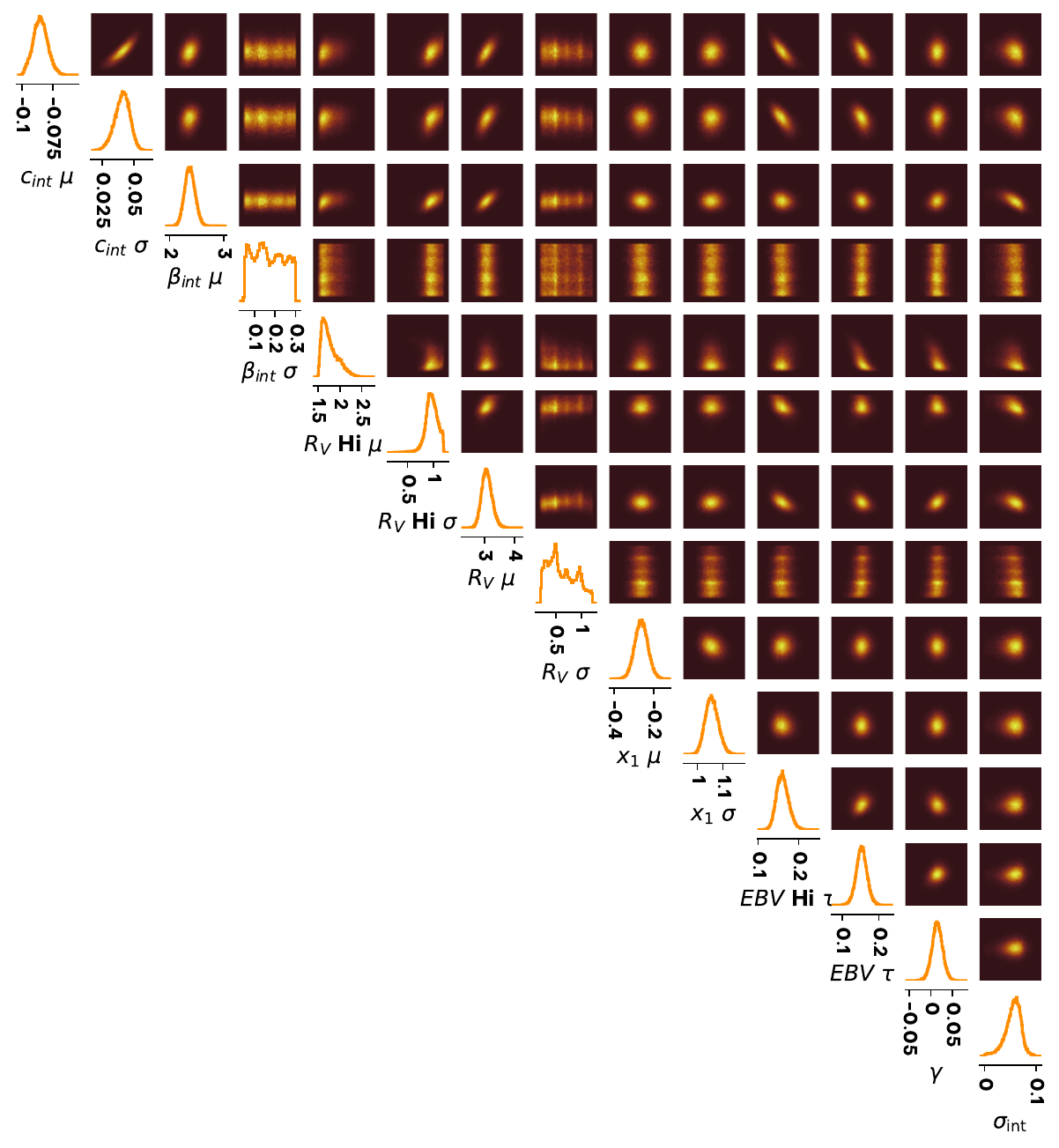}
    \caption{Posteriors from the nominal Stjörnumál model, on the DES-Dovekie data. Orange histograms show the 1D posteriors, and correlations are shown for each parameter. In general, the mean values are well-constrained, though some of the $\sigma$ values do not converge. }
    \label{fig:posteriors}
\end{figure*}

Figure \ref{fig:posteriors} shows the corner plot for the Stjörnumál posterior. In general, we find our posteriors to be well-constrained, with the exception of $\beta_{\rm int} \sigma$. We find no meaningful constraints on $\sigma_{\beta_{SN}}$, though we can see from the PPC (Section \ref{sec:Validation:subsubsec:PPC}) that this does not impact our simulations. Table \ref{tab:results} shows the Stjörnumál output for each of our models.

\def\NOMcintmu{$-0.085(0.01)$}
\def\NOMcintsigma{$0.041(0.01)$}
\def\NOMbetaintmu{$2.371(0.09)$}
\def\NOMbetaintsigma{$0.171(0.07)$}
\def\NOMRVHimu{$1.769(0.18)$}
\def\NOMRVHisigma{$0.965(0.12)$}
\def\NOMRVmu{$3.081(0.17)$}
\def\NOMRVsigma{$0.618(0.26)$}
\def\NOMxmu{$-0.264(0.03)$}
\def\NOMxsigma{$1.057(0.03)$}
\def\NOMEBVHitau{$0.161(0.01)$}
\def\NOMEBVtau{$0.153(0.01)$}
\def\NOMgamma{$0.015(0.01)$}
\def\NOMsigmarmint{$0.055(0.02)$}
\def\LOGcintmu{$-0.061(0.02)$}
\def\LOGcintsigma{$0.055(0.01)$}
\def\LOGbetaintmu{$2.477(0.21)$}
\def\LOGbetaintsigma{$0.187(0.07)$}
\def\LOGxmu{$-0.301(0.04)$}
\def\LOGxsigma{$1.076(0.03)$}
\def\LOGEBVHitau{$0.140(0.03)$}
\def\LOGEBVtau{$0.081(0.02)$}
\def\LOGRVL{$1.447(0.41)$}
\def\LOGRVk{$0.368(0.83)$}
\def\LOGRVsigma{$0.290(0.13)$}
\def\LOGgamma{$-0.002(0.00)$}
\def\LOGsigmarmint{$0.051(0.03)$}
\def\TWOCOLcintHimu{$-0.075(0.01)$}
\def\TWOCOLcintHisigma{$0.047(0.01)$}
\def\TWOCOLcintmu{$-0.092(0.00)$}
\def\TWOCOLcintsigma{$0.033(0.01)$}
\def\TWOCOLRVmu{$2.322(0.20)$}
\def\TWOCOLRVsigma{$1.042(0.18)$}
\def\TWOCOLbetaintHimu{$2.609(0.16)$}
\def\TWOCOLbetaintHisigma{$0.216(0.07)$}
\def\TWOCOLbetaintmu{$1.999(0.11)$}
\def\TWOCOLbetaintsigma{$0.165(0.07)$}
\def\TWOCOLxmu{$-0.250(0.03)$}
\def\TWOCOLxsigma{$1.024(0.02)$}
\def\TWOCOLEBVtau{$0.150(0.01)$}
\def\TWOCOLsigmarmint{$0.081(0.01)$}
\def\MAScintmu{$-0.081(0.00)$}
\def\MAScintsigma{$0.044(0.00)$}
\def\MASbetaintmu{$2.222(0.07)$}
\def\MASbetaintsigma{$0.179(0.07)$}
\def\MASRVHimu{$1.716(0.13)$}
\def\MASRVHisigma{$0.939(0.11)$}
\def\MASRVmu{$2.916(0.15)$}
\def\MASRVsigma{$0.680(0.28)$}
\def\MASxmu{$-0.239(0.03)$}
\def\MASxsigma{$1.052(0.02)$}
\def\MASEBVHitau{$0.157(0.01)$}
\def\MASEBVtau{$0.151(0.01)$}
\def\MASsigmarmint{$0.089(0.01)$}

\def\BANcintmuA{$-0.080(0.01)$}
\def\BANcintsigmaA{$0.042(0.01)$}
\def\BANRVmuA{$2.170(0.33)$}
\def\BANRVsigmaA{$0.984(0.28)$}
\def\BANbetaintmuA{$2.518(0.15)$}
\def\BANbetaintsigmaA{$0.180(0.07)$}
\def\BANxmuA{$0.171(0.09)$}
\def\BANxsigmaA{$0.741(0.07)$}
\def\BANEBVtauA{$0.144(0.02)$}
\def\BANcintmuB{$-0.073(0.02)$}
\def\BANcintsigmaB{$0.051(0.02)$}
\def\BANRVmuB{$3.082(0.73)$}
\def\BANRVsigmaB{$0.983(0.28)$}
\def\BANxmuB{$-1.668(0.35)$}
\def\BANxsigmaB{$0.719(0.28)$}
\def\BANEBVtauB{$0.118(0.04)$}
\def\BANfmathrmmix{$0.781(0.10)$}
\def\BANgamma{$0.079(0.04)$}
\def\BANsigmarmint{$0.079(0.01)$}

\def\LBNcintAmu{$-0.067(0.02)$}
\def\LBNcintAsigma{$0.044(0.01)$}
\def\LBNRVmu{$2.294(0.26)$}
\def\LBNRVsigma{$1.063(0.19)$}
\def\LBNbetaintmu{$2.259(0.14)$}
\def\LBNbetaintsigma{$0.180(0.07)$}
\def\LBNxAmu{$0.274(0.15)$}
\def\LBNxAsigma{$0.587(0.19)$}
\def\LBNtau{$0.149(0.01)$}
\def\LBNcintBmu{$-0.086(0.01)$}
\def\LBNcintBsigma{$0.049(0.01)$}
\def\LBNxBmu{$-0.877(0.46)$}
\def\LBNxBsigma{$1.069(0.14)$}
\def\LBNfmathrmmix{$0.476(0.23)$}
\def\LBNgamma{$0.018(0.04)$}
\def\LBNsigmarmint{$0.075(0.01)$}

\def\SRVcintmu{$-0.080(0.01)$}
\def\SRVcintsigma{$0.045(0.00)$}
\def\SRVbetaintmu{$2.172(0.07)$}
\def\SRVbetaintsigma{$0.176(0.07)$}
\def\SRVRVmu{$2.257(0.21)$}
\def\SRVRVsigma{$1.015(0.13)$}
\def\SRVxmu{$-0.269(0.03)$}
\def\SRVxsigma{$1.025(0.03)$}
\def\SRVEBVHitau{$0.174(0.01)$}
\def\SRVEBVtau{$0.105(0.01)$}
\def\SRVgamma{$-0.044(0.01)$}
\def\SRVsigmarmint{$0.084(0.01)$}

\clearpage
\vspace*{-1cm}
\hspace*{2cm}
\begin{sidewaystable}
\centering

\resizebox{\textheight}{!}{%
\begin{tabular}{l|llllllllllllll}
Parameter                   & \multicolumn{2}{c}{\textbf{Nominal}}                      & \multicolumn{2}{c}{\textbf{No Mass Step}}               & \multicolumn{2}{c}{\textbf{Logistic $R_V$}}                   & \multicolumn{2}{c}{\textbf{Single Rv}}                  & \multicolumn{2}{c}{\textbf{2 Colours}}                                      & \multicolumn{2}{c}{\textbf{Banana Split}}                           & \multicolumn{2}{c}{\textbf{Banana Less Split}}                      \\
\hline
$\overline{c}_{\rm int}$    & \multicolumn{2}{c}{\NOMcintmu}             & \multicolumn{2}{c}{\MAScintmu}           & \multicolumn{2}{c}{\LOGcintmu}           & \multicolumn{2}{c}{\SRVcintmu}           & \TWOCOLcintmu         & \TWOCOLcintHimu       & \BANcintmuA       & \BANcintmuB       & \LBNcintAmu       & \LBNcintBmu       \\
$\sigma_{c_{\rm int}}$      & \multicolumn{2}{c}{\NOMcintsigma}          & \multicolumn{2}{c}{\MAScintsigma}        & \multicolumn{2}{c}{\LOGcintsigma}        & \multicolumn{2}{c}{\SRVcintsigma}        & \TWOCOLcintsigma      & \TWOCOLcintHisigma    & \BANcintsigmaA    & \BANcintsigmaB    & \LBNcintAsigma    & \LBNcintBsigma    \\
$\overline{R}_V$            & \NOMRVmu    & \NOMRVHimu   & \MASRVmu   & \MASRVHimu   & \multicolumn{2}{c}{N/A}                    & \multicolumn{2}{c}{\SRVRVmu}              & \multicolumn{2}{c}{\TWOCOLRVmu}                              & \BANRVmuA         & \BANRVmuB         & \multicolumn{2}{c}{\LBNRVmu}                         \\
$\sigma_{R_V}$              & \NOMRVsigma & \NOMRVHisigma & \MASRVsigma   & \MASRVHisigma   & \multicolumn{2}{c}{\LOGRVsigma}          & \multicolumn{2}{c}{\SRVRVsigma}           & \multicolumn{2}{c}{\TWOCOLRVsigma}                           & \BANRVsigmaA      & \BANRVsigmaB      & \multicolumn{2}{c}{\LBNRVsigma}                      \\
$\overline{\beta}_{\rm SN}$ & \multicolumn{2}{c}{\NOMbetaintmu}          & \multicolumn{2}{c}{\MASbetaintmu}        & \multicolumn{2}{c}{\LOGbetaintmu}        & \multicolumn{2}{c}{\SRVbetaintmu}         & \TWOCOLbetaintmu      & \TWOCOLbetaintHimu    & \multicolumn{2}{c}{\BANbetaintmuA}   & \multicolumn{2}{c}{\LBNbetaintmu}   \\
$\sigma_{\beta_{\rm int}}$   & \multicolumn{2}{c}{\NOMbetaintsigma}       & \multicolumn{2}{c}{\MASbetaintsigma}     & \multicolumn{2}{c}{\LOGbetaintsigma}     & \multicolumn{2}{c}{\SRVbetaintsigma}      & \TWOCOLbetaintHisigma & \TWOCOLbetaintHisigma & \multicolumn{2}{c}{\BANbetaintsigmaA} & \multicolumn{2}{c}{\LBNbetaintmu} \\
$\overline{x}_1$            & \multicolumn{2}{c}{\NOMxmu}                & \multicolumn{2}{c}{\MASxmu}              & \multicolumn{2}{c}{\LOGxmu}              & \multicolumn{2}{c}{\SRVxmu}                & \multicolumn{2}{c}{\TWOCOLxmu}                               & \BANxmuA          & \BANxmuB          & \LBNxAmu          & \LBNxBmu          \\
$\sigma_{x_1}$              & \multicolumn{2}{c}{\NOMxsigma}             & \multicolumn{2}{c}{\MASxsigma}           & \multicolumn{2}{c}{\LOGxsigma}           & \multicolumn{2}{c}{\SRVxsigma}             & \multicolumn{2}{c}{\TWOCOLxsigma}                            & \BANxsigmaA       & \BANxsigmaB       & \LBNxAsigma       & \LBNxBsigma       \\
$\tau_{\rm E}$              & \NOMEBVtau  & \NOMEBVHitau  & \MASEBVtau & \LOGEBVHitau & \LOGEBVtau & \LOGEBVHitau & \multicolumn{2}{c}{\SRVEBVtau}            & \multicolumn{2}{c}{\TWOCOLEBVtau}                            & \BANEBVtauA          & \BANEBVtauB          & \multicolumn{2}{c}{\LBNtau}                          \\
$\gamma$                    & \multicolumn{2}{c}{\NOMgamma}              & \multicolumn{2}{c}{--}                    & \multicolumn{2}{c}{\LOGgamma}            & \multicolumn{2}{c}{\SRVgamma}              & \multicolumn{2}{c}{--}                                                      & \multicolumn{2}{c}{\BANgamma}                                              & \multicolumn{2}{c}{\LBNgamma}                                              \\
$\sigma_{\rm int}$          & \multicolumn{2}{c}{\NOMsigmarmint}         & \multicolumn{2}{c}{\MASsigmarmint}       & \multicolumn{2}{c}{\LOGsigmarmint}       & \multicolumn{2}{c}{\SRVsigmarmint}         & \multicolumn{2}{c}{\TWOCOLsigmarmint}                        & \multicolumn{2}{c}{\BANsigmarmint}   & \multicolumn{2}{c}{\LBNsigmarmint}  \\
\hline
$\log(\mathcal{Z})$         & \multicolumn{2}{c}{--}                     & \multicolumn{2}{c}{$-0.232 (0.06)$}                 & \multicolumn{2}{c}{0.008 (0.002)}                & \multicolumn{2}{c}{0.010(0.003)}                   & \multicolumn{2}{c}{0.169(0.366)}                                                    & \multicolumn{2}{c}{1.9(0.77)}                                            & \multicolumn{2}{c}{3.4(0.8) }       \\
$\chi^2/\nu$         & \multicolumn{2}{c}{36.5}                     & \multicolumn{2}{c}{53.7}                 & \multicolumn{2}{c}{37.8}                & \multicolumn{2}{c}{64.7}                   & \multicolumn{2}{c}{48.5}                                                    & \multicolumn{2}{c}{51.9}                                            & \multicolumn{2}{c}{55.9}  
\end{tabular}
}
\caption{Mean posterior values and their associated uncertainties for each tested model. We provide the $\log(\mathcal{Z})$ and $\chi^2/\nu$ value for each model as well.}
\label{tab:results}
\end{sidewaystable}
\clearpage


\subsection{Model Preference}\label{sec:Results:subsec:Model}

Following the evidences set out by \cite{Trotta08}, we find moderate-to-weak evidence against two populations of SNe Ia in the DES data set (Banana Split $\log(10)\textrm{BF} = 1.90$, Banana Less Split $\log(10)\textrm{BF} = 3.4$ ). For these two mixture models, we find a mixture fraction $f_{\rm mix}$ of \BANfmathrmmix \, for Banana Split and \LBNfmathrmmix \, for Banana Less Split, respectively. These numbers align well with the high-redshift values from \cite{Rubin26}, and indicate a predominantly single population at high redshift. When comparing the  $\log(\mathcal{Z})$ with the \texttt{Dust2Dust} model metrics\footnote{Equations 7-14 in \cite{Popovic22}} - the difference between sims and data for the $c$ histogram, binned $c$ vs $\mu_{\rm RES}$, and the scatter of $\mu_{\rm RES}$ as a function of $c$, the story is much the same: Both Bananas Split perform worse than the nominal model, but not worse than any other model. Single $R_V$ is the worst-performing model at $\chi^2/\nu \sim 65$, twice as bad as the nominal model and $\Delta \chi^2/\nu \sim 10$ worse than No Mass Step, 2 Colours, and the Bananas Split. The Logistic model performs equally-well to the Nominal model, with $L = 1.45 \pm 0.41$, $k = 0.36 \pm 0.083$, $\sigma = 0.29 \pm 0.13$. The worse $\chi^2/\nu$ performance of the `No Mass Step' model is driven by the increased $\sigma_{\rm int} = 0.089$ as compared to the nominal $\sigma_{\rm int} = 0.055$. Reducing this $\sigma_{\rm int}$ and re-simulating does improve the $\chi^2/\nu$ and does not appear to significantly impact the PPC; nonetheless, the model prefers this higher $\sigma_{\rm int}$.

Put together, the frequentist and Bayesian approaches indicate that while the Banana Split model is well-enough able to replicate the data, the additional model parameters do not add a sufficient amount of fidelity for the data to prefer Banana Split. 

\begin{table}
    \centering
    \begin{tabular}{c|ccc}
    Model &
    $c$ Distribution &
    $\mu - \mu_{\rm model}$ &
    $\sigma_{\mu - \mu_{\rm model}}$ \\
                  &
    $\chi^2/\nu$  &
    $\chi^2/\nu$  &
    $\chi^2/\nu$  \\
    \hline
    \textbf{Nominal} & 1.83 & 9.28 & 26.13 \\
    \textbf{No Mass Step} & 2.87 & 9.45 & 41.40 \\
    \textbf{Logistic $R_V$} & 2.00 & 11.98 & 25.84 \\
    \textbf{Single Rv} & 2.41 & 17.9 & 40.81 \\
    \textbf{2 Colours} & 1.77 & 7.19 & 37.43 \\
    \textbf{Banana Split} & 1.29 & 18.51 & 32.14 \\
    \textbf{Banana Less Split} & 3.64 & 18.24 & 34.01 \\
    \end{tabular}
    \caption{A breakdown of the total \texttt{Dust2Dust} metric value shown in Table \ref{tab:results}.}
    \label{tab:D2DMetric}
\end{table}

It is worth pointing out that the $\Delta \chi^2/\nu$ dust model metrics do \textit{not} include a $x_1$ term; the dust models make no direct prediction on $x_1$ distributions. The Bayes' factor \textit{does} include the $x_1$ distribution alongside everything else, and represents a more holistic evaluation of the model. Table \ref{tab:D2DMetric} shows the specific breakdown of $\chi2/\nu$ for each of the \texttt{D2D} metrics for each model; we include visualisations in Appendix \ref{sec:appendix}. In all cases, we use the same priors as shown in Table \ref{tab:priors}.

\subsection{On The Observed $\Delta R_V$}\label{sec:Results:subsec:MASS}

\begin{figure}
    \centering
    \includegraphics[width=8cm]{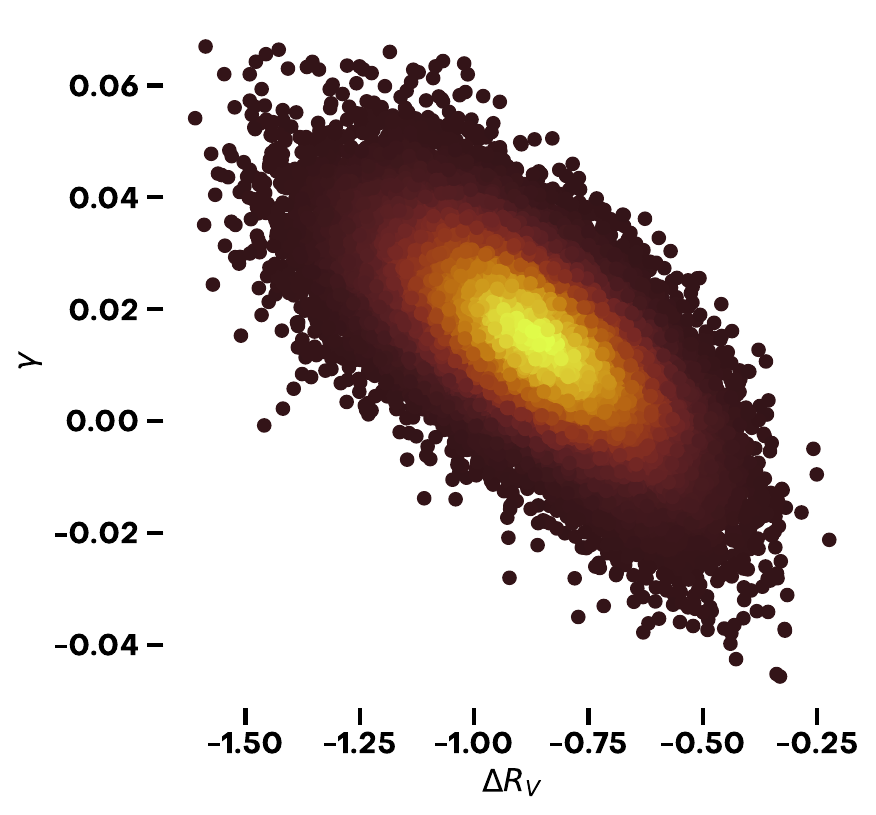}
    \caption{Inferred $\Delta\mathop{\mathbb{E}}(R_V)$ vs $\gamma$. Points are taken from the posterior draws of Figure \ref{fig:posteriors} and colour-coded by density. For $\Delta R_V$ we use the truncated normal moment as shown in Section \ref{sec:Results:subsec:MASS}}
    \label{fig:GAMMA}
\end{figure}

Our high mass $R_V~\mu$ distribution is truncated at $R_V = 1.2$. We calculate the expectation value for $R_V$ with
$$\alpha = (S - \mu)/\sigma$$
$$\phi = (1/\sqrt{2\pi})*\exp(-\alpha^2/2)$$
$$Z = 0.5 (1- \textrm{erf}(\alpha/\sqrt{2}))$$
\begin{equation}
    \mathop{\mathbb{E}}(\mu) = (\mu + \sigma)\times  \frac{\phi}{Z}
\end{equation}
where $S$ is the lower $R_V$ limit, here $S=1.2$. Our expected $R_V~\mu$ values for high and low mass are $2.21$ and $3.08$, respectively, a $\Delta\mathop{\mathbb{E}}(R_V) = 0.8$. We show the posterior distribution of $\Delta\mathop{\mathbb{E}}(R_V)$ and the mass step $\gamma$ in Figure \ref{fig:GAMMA}, where we see that a $\gamma = 0 $ corresponds to $\Delta\mathop{\mathbb{E}}(R_V) = 0.6$. The Logistic model gives a $\Delta \overline{R}_V = 0.44$, with an overall $\overline{R}_V = 0.44$.

The remaining key difference between Stjörnumál and BayeSN fits is that BayeSN fits population-level $R_V$ values as part of the model-training process, in contrast to the limitations of SALT. Stjörnumál relies on catalogue-level SALT fits, where the SALT training and lightcurve fitting are already performed. 

\subsection{Impact on Distances}

We refrain from directly computing $\Delta w$ from the Tripp distances. \cite{Popovic24c} shows that comparing un-bias corrected distances can bias $\Delta w$ values\footnote{Specifically, the first iteration found a calibration-induced $\Delta w = 0.10$, a $\times 5$ increase from the eventual full-cosmology $\Delta w = 0.02$.}. Previous attempts at demonstrating $\Delta w$ with uncorrected distances assumed that the selection effects and biases would be the similar enough across tested systematics to effectively cancel each other out. Evidently, this is not so: \cite{Popovic26} performed an end-to-end cosmology analysis and found the resulting $\Delta w$ from changes in calibration were only $\Delta w =0.02$. 

\begin{figure}
    \centering
    \includegraphics[width=8cm]{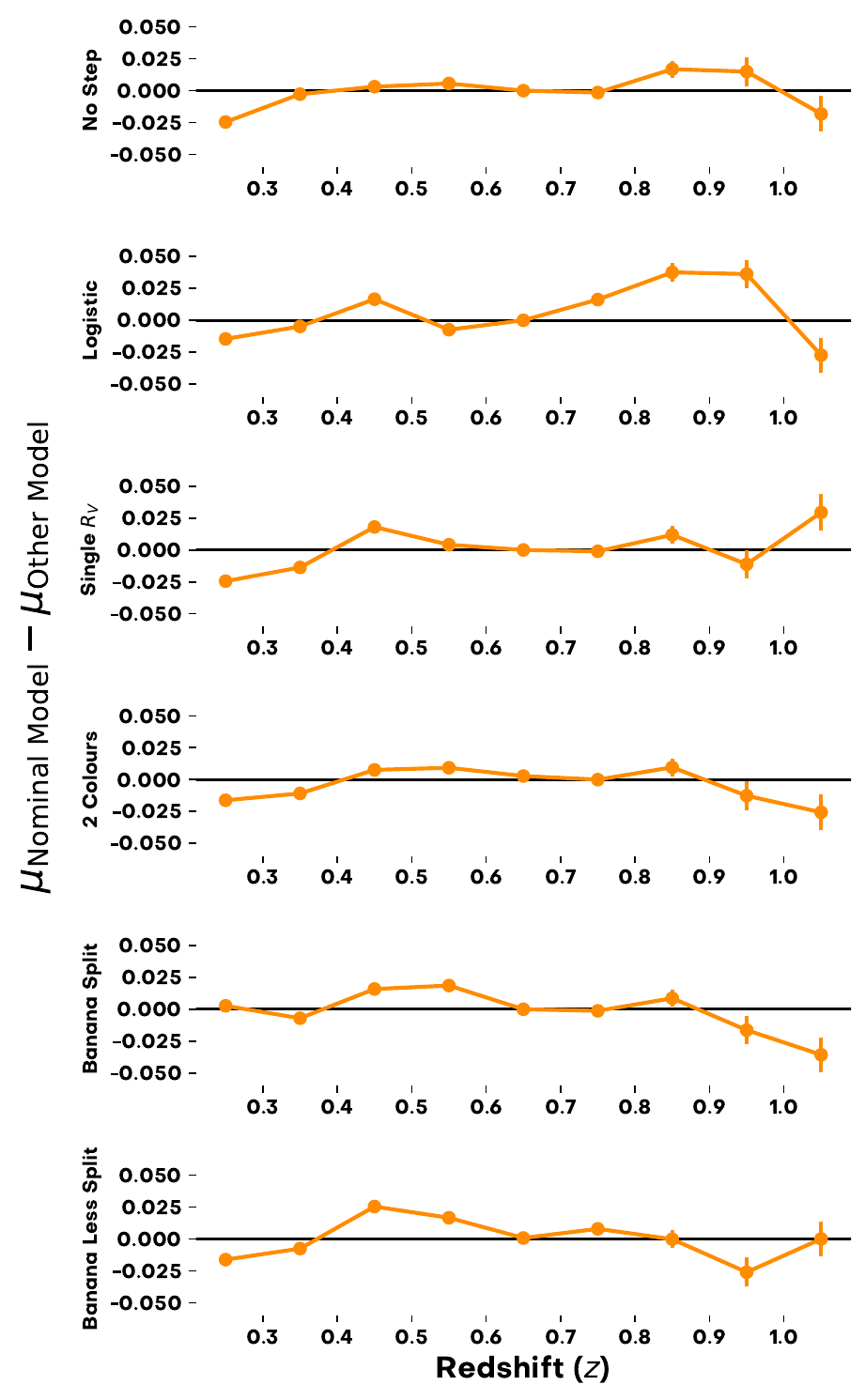}
    \caption{$\Delta \mu$ between the \textbf{Nominal} model employed in this paper and each other tested model. Orange points show the difference in median $\mu$ in each bin, black line shows $\Delta \mu =0$. We see $d\mu/dz < 0.025$ in all cases.}
    \label{fig:dmudz}
\end{figure}

Instead, we elect to show $d\mu/dz$ between the various (simulated) models. Figure \ref{fig:dmudz} shows the difference in median $\mu$ values as a function of redshift, between the \textbf{Nominal} model and each of the other tested models. This is done for $50\times$ the size of the DES5YR data set; we see relatively little redshift-dependent behaviour with a max $d\mu/dz$ of 0.025.

\section{Discussion \& Conclusions}\label{sec:Discussion}

In this work, we have presented a novel simulation based inference approach to inferring intrinsic SNe Ia and extrinsic dust populations, providing more flexibility and significant speed upgrades to the previous cosmology pipeline-integrated \texttt{Dust2Dust}. We test the SBI framework on several models, and with the addition of a model classifier, we are able to directly compare potential models of SNe Ia. We find that a combination of Bayes' Factor and frequentist statistics point slightly towards two populations of SNe Ia traced by $x_1$, and that a logistically-distributed $R_V$ distribution performs better in the previously-set out metrics that describe the dust models introduced by \cite{BS20}. 

We also find that the Stjörnumál pipeline provides well-calibrated errors from Simulation Based Calibration and Tests of Accuracy with Random Points, and that these results are robust to $|\Delta w| < 0.1$.

\subsection{On The Origin of the Mass Step}
Previous mass step analyses using BayeSN \citep{TM22, Grayling24, GP25, Ginolin26} have found $\Delta\mathop{\mathbb{E}}(R_V) $ values ranging from $\sim0.1-0.5$ when allowing for simultaneously differing $R_V$ distributions and an achromatic magnitude difference, which they typically find to be $\sim 0.05$ mag. \citet{Ginolin26} finds a larger achromatic difference of $\sim0.1$ mag on the ZTF DR2 sample, though this large environmental dependence seems to be a characteristic feature of that data set as established in other work \citep[][]{Ginolin24b, Ginolin25}.

Our results show some disagreement with these. When allowing for an achromatic mass step, we find $\gamma = $ \NOMgamma, consistent with 0 to within $2\sigma$. However, this result is also not in strong tension with the result from \citet{Grayling24} of $0.049\pm0.016$ mag; combining the uncertainties in quadrature, the discrepancy between the two results is $\sim1.8\sigma$. Given the methodological differences, this mild disagreement is not surprising. We find a $\gamma$ more consistent with theirs only when barring a mass-dependent $R_V$, though this mass-independent $\overline{R}_V$ is consistent with their recovered $R_V$.

Previous \texttt{Dust2Dust} results differed from the approach in \cite{Grayling24} in three primary respects outside of the choice in post-SALT inference (\texttt{Dust2Dust)} and model training in BayeSN:

\begin{itemize}
    \item \textbf{Explicit Achromatic Mass Step:} The original \texttt{Dust2Dust} code did not allow for an achromatic mass step, unlike BayeSN.
    \item \textbf{Lower $R_V$ Limit:} \cite{Grayling24} placed a lower $R_V = 1.2$ limit on their model, unlike \texttt{Dust2Dust} which enforced the less-realistic $R_V >0.4$.
    \item \textbf{Expectation Value of $R_V$:} \cite{Grayling24} provided expectation values of their $R_V$ values to explicitly account for the truncated distribution, not performed in \texttt{Dust2Dust}.
\end{itemize}

In this analysis, we have explicitly addressed these three issues; the discrepant $\Delta \overline{R}_V$ remains. Instead, one of two conclusions presents itself. As seen in Table \ref{tab:results}, the two-population `Banana Split' model performs equally-well to the Nominal model, with a smaller $\Delta \overline{R}_V$, and is well-evidenced by the literature \citep{Wojtak23, Rubin26}. The other alternative model is the Logistic model, with a comparable $\Delta \overline{R}_V = 0.44$ comparable to that of \cite{Grayling24}. Neither of these models provide a direct solution to the divergent $\Delta \overline{R}_V$, but do work towards a more consistent $\Delta \overline{R}_V$ and a more realistic overall $\overline{R}_V$ \citep{Murakami25}.

\subsection{SN Populations}

We find evidence against two populations of SNe Ia at high redshift. Given ongoing studies linking progenitors to differing stretch distributions \citep{Rigault2020, Wojtak26, Wiseman22}, this is not terribly surprising: the high-redshift stretch distribution is well-described by a single Gaussian. 

We stress that, despite Banana Split receiving a $\log(10)\textrm{BF} = 1.9$, this does not make it a bad model. The original intent of Banana Split is to describe the entire redshift range, rather than just high-redshift. While the Banana Split model does well-describe the data at high redshift, the additional $\times 2$ number of parameters penalises the model over the dust-based approaches. 

\subsection{Model Identification}

In Sections \ref{sec:Validation:subsec:Miss} and \ref{sec:Results:subsec:Model}, we see that the current quality and quantity of data is insufficient to differentiate between differing ways of distributing $R_V$. This holds even within the internal Stjörnumál simulations for the \textbf{Single $R_V$} and \textbf{Logistic $R_V$}, though interestingly we do find weak discernment between \textbf{No Mass Step} and the \textbf{Nominal} model. We are well-able to identify two-population models. In either case, more data is needed to switch entirely to the Bayes Factor as a means of identification, or a more complex weighting than conditioning the neural network solely on the raw distributions. For instance, \textbf{Single $R_V$} returns a $\chi^2/\nu = 64.7$, a $\times2$ increase from the \textbf{Nominal} model; this is not reflected in the Bayes Factor, which finds no conclusive difference between the two. For now, it is clear that a combination of the Bayes Factor and the more conventional metrics is the best path forwards. 

\subsection{Future Work}

There remain several shortcomings to Stjörnumál, most notably the reliance on the simulation bank. This simulation bank precludes the calculation of nuisance parameters $\alpha$ and $\beta$, and explicitly disallows for Kronecker-delta functions. Incorporating more rapid and pythonic simulation software, such as \texttt{skysurvey} \citep{Rigault26}, will ameliorate the above issues, and also allow for varying cosmology simultaneously with the SN and dust parameters. 

Further improvements relate to calculating $\gamma$; currently, Stjörnumál calculates $\gamma$ with an ad-hoc addition to the $\mu_{\rm RES}$ and $m_B$ distributions. Studies such as \cite{Ginolin24b} have argued that this is inappropriate and that $\gamma$ should be calculated alongside $\alpha_{\rm SALT}$ and $\beta_{\rm SALT}$. While Stjörnumál does possess this ability, performing an MCMC fit on each simulation is currently prohibitively expensive: a single MCMC run takes $O(10)$seconds, and operates on a different parallelisation scheme than the simulations. Taken together, this would change the simulation time from $O$minutes to $O$100s of hours. A more clever coding situation is needed. 

Here we have run Stjörnumál solely on high-redshift data, but the infrastructure allows for redshift-dependent parameter evolution. Combining Stjörnumál with high-stats low-redshift samples, such as DEBASS \citep{Acevedo26} or TITAN (Tweddle et al., \textit{in prep.}) will enable further exploration of these effects, and a more thorough evaluation of Banana Split. 

Finally, recent work by \cite{Shruti25} and others indicates that high mass star-forming and quiescent galaxies produce different $\beta$ values for SNe Ia, and may point to different dust or intrinsic distributions. Stjörnumál presents a unique opportunity to explore this new parameter space.

\section*{Acknowledgments}

This project has received funding from the European Union’s Horizon Europe research and innovation programme under the Marie Skłodowska-Curie grant agreement No 101205780.

Generative AI was used in the development of this paper; specifically for GPU integration and the implementation of the mixture model. 

B.P thanks R. Wojtak for discussions about the project. B.P would like to acknowledge the reader for taking the time to read this paper, as well as Millie Smith for her work on the Stjörnumál logo on the github. 

\section{Data Availability}

The DES-Dovekie distances and light curves are available at \url{https://github.com/des-science/DES-SN5YR}, and we provide the github at \url{https://github.com/bap37/Stjornumal}.


\appendix

\section{\texttt{Dust2Dust} Metrics}\label{sec:appendix}

Here we present the \texttt{Dust2Dust} metrics for each of the tested models. 

\begin{figure*}
    \centering
    \includegraphics[width=18cm]{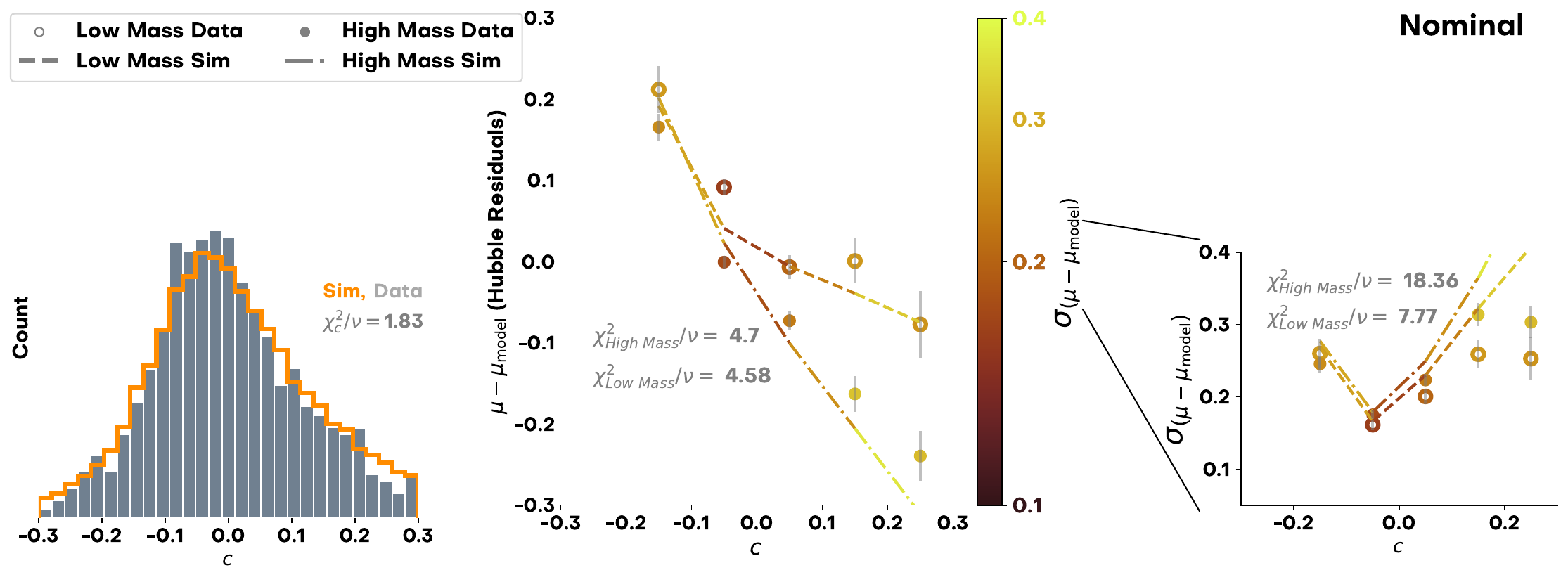}
    \caption{\texttt{Dust2Dust} metrics for the Nominal model. }
    \label{fig:NOM_METRIC}
\end{figure*}

\begin{figure*}
    \centering
    \includegraphics[width=18cm]{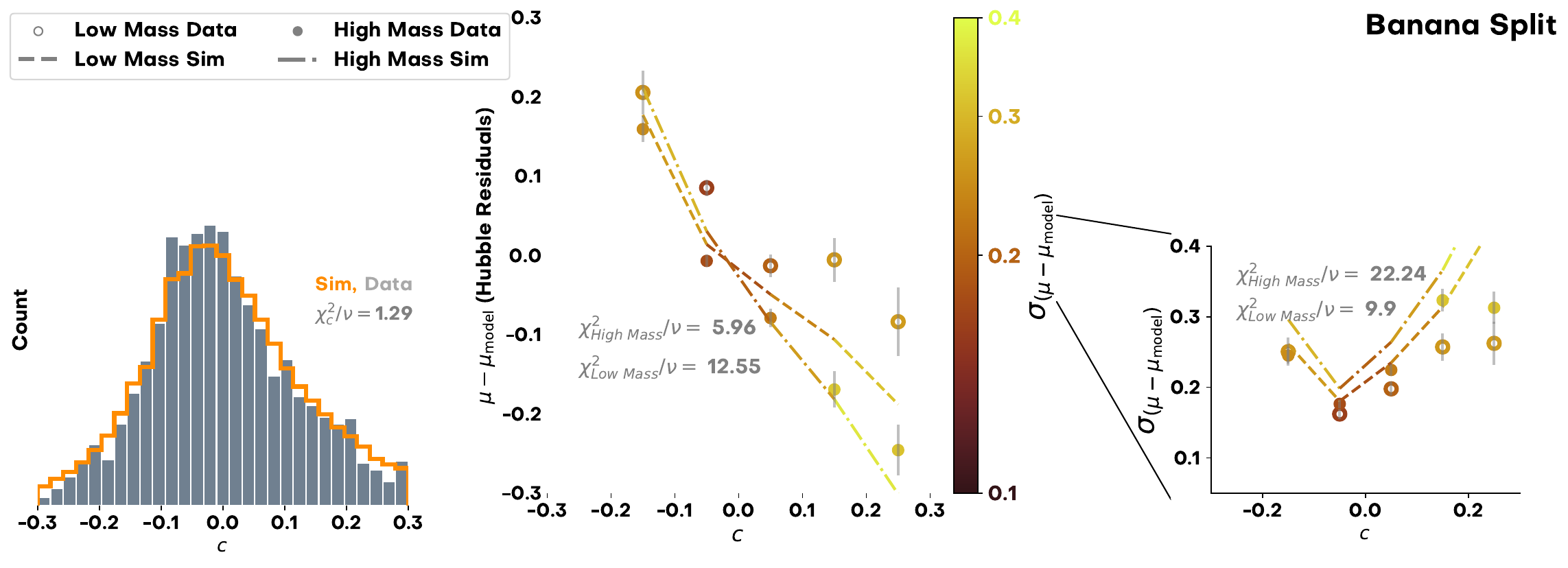}
    \caption{\texttt{Dust2Dust} metrics for the Banana Split model. }
    \label{fig:BAN_METRIC}
\end{figure*}

\begin{figure*}
    \centering
    \includegraphics[width=18cm]{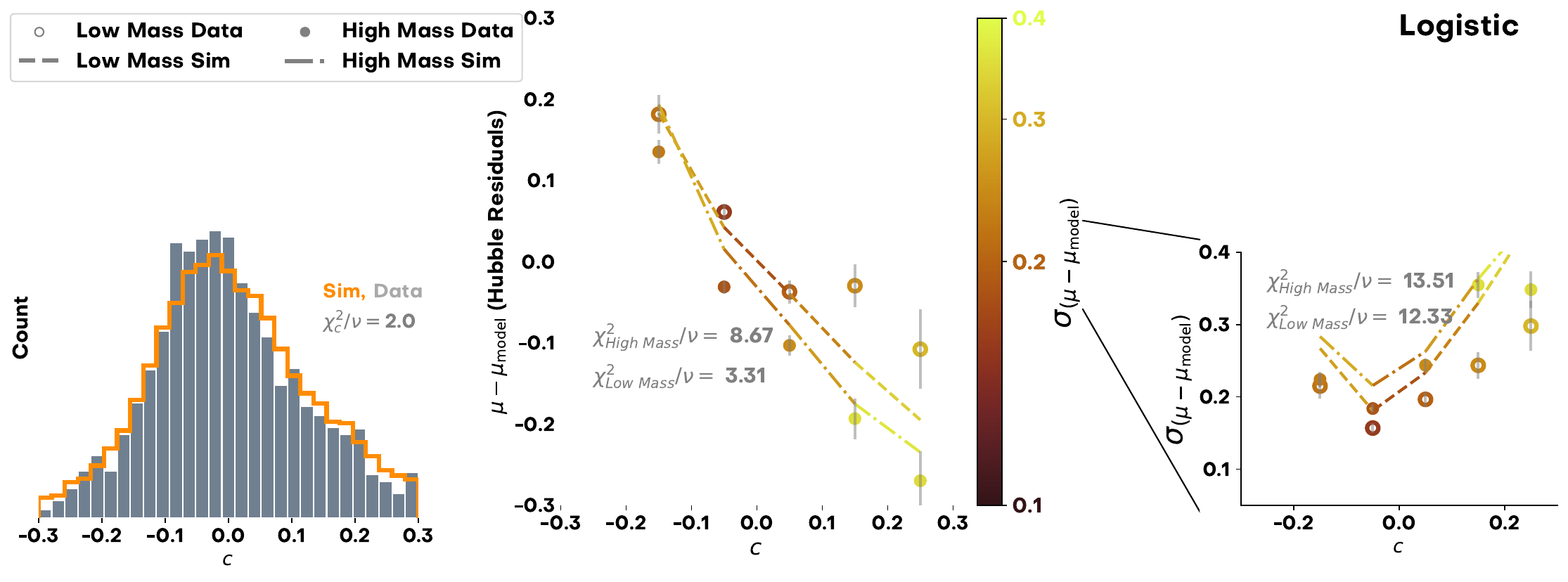}
    \caption{\texttt{Dust2Dust} metrics for the Logistic model. }
    \label{fig:LOG_METRIC}
\end{figure*}

\begin{figure*}
    \centering
    \includegraphics[width=18cm]{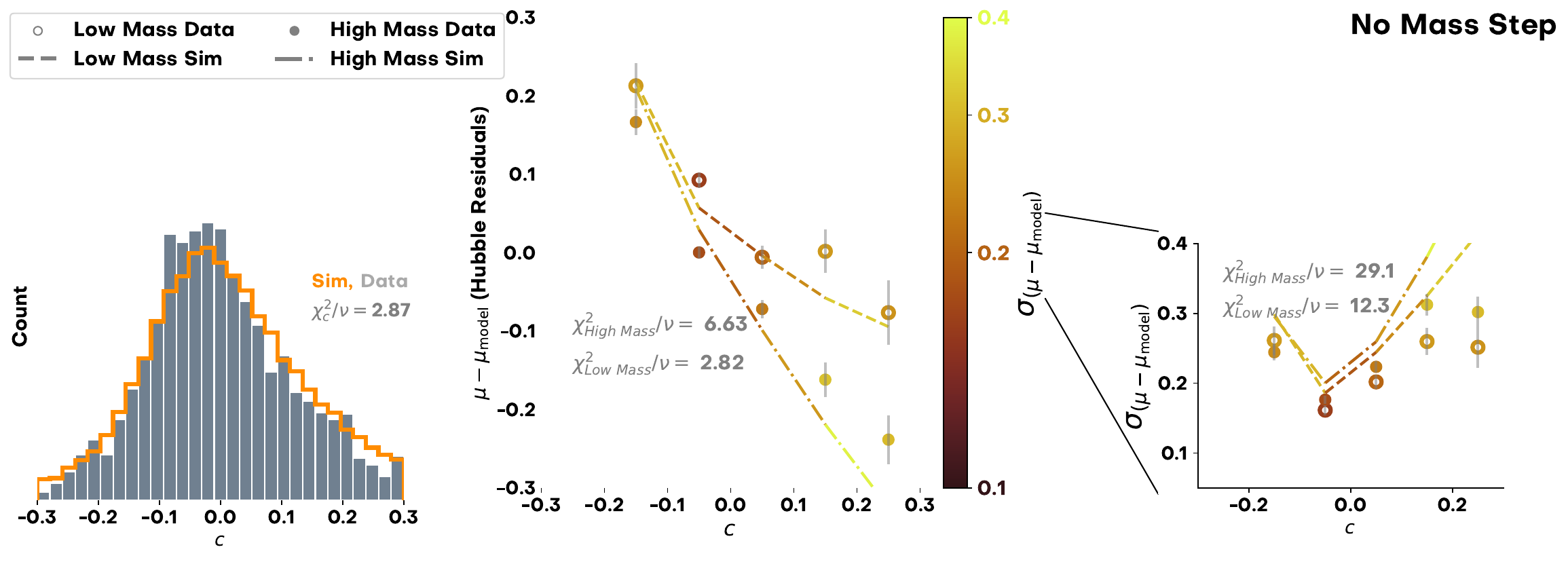}
    \caption{\texttt{Dust2Dust} metrics for the No Mass Step model. }
    \label{fig:NOS_METRIC}
\end{figure*}

\begin{figure*}
    \centering
    \includegraphics[width=18cm]{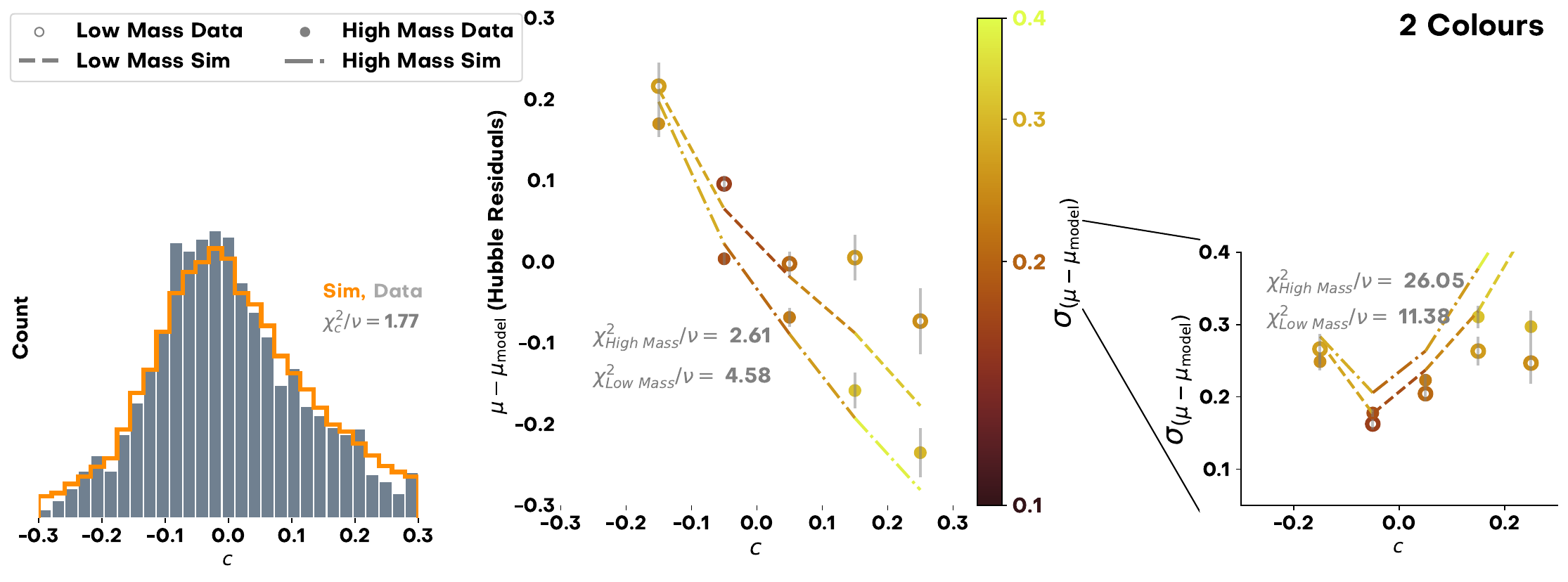}
    \caption{\texttt{Dust2Dust} metrics for the 2 Colours model. }
    \label{fig:2COL_METRIC}
\end{figure*}

\begin{figure*}
    \centering
    \includegraphics[width=18cm]{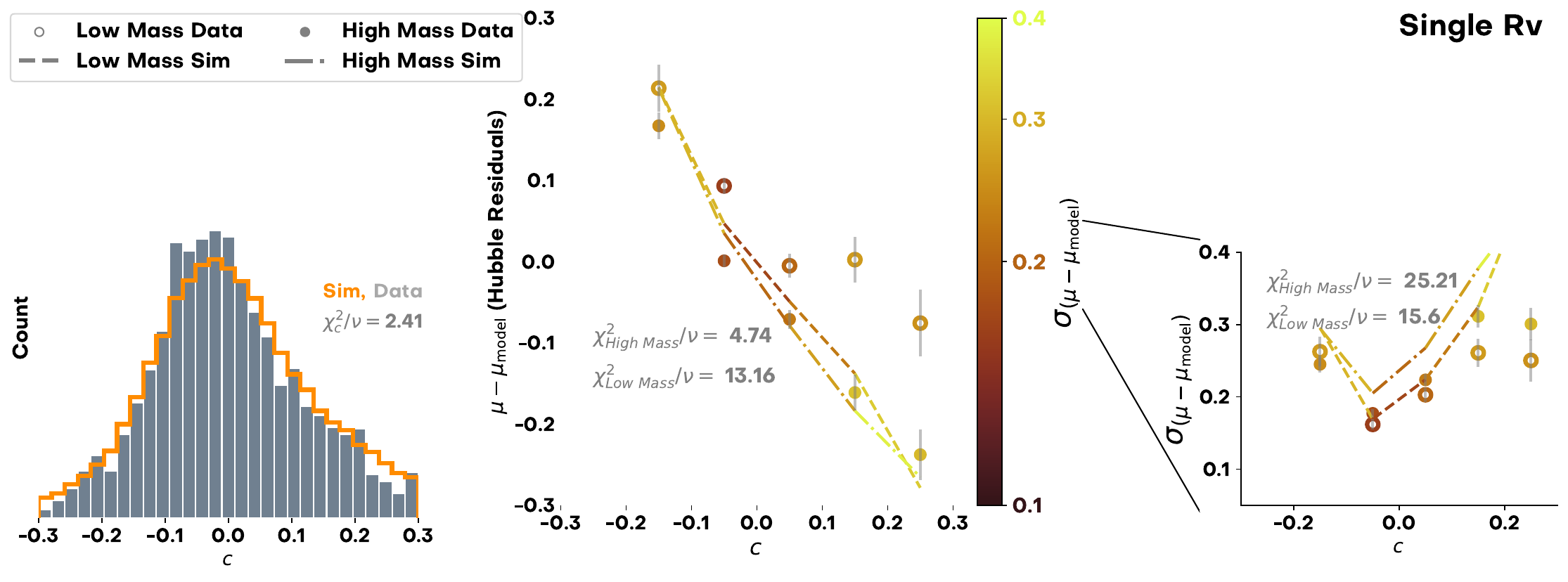}
    \caption{\texttt{Dust2Dust} metrics for the Single $R_V$ model. }
    \label{fig:SRV_METRIC}
\end{figure*}

\begin{figure*}
    \centering
    \includegraphics[width=18cm]{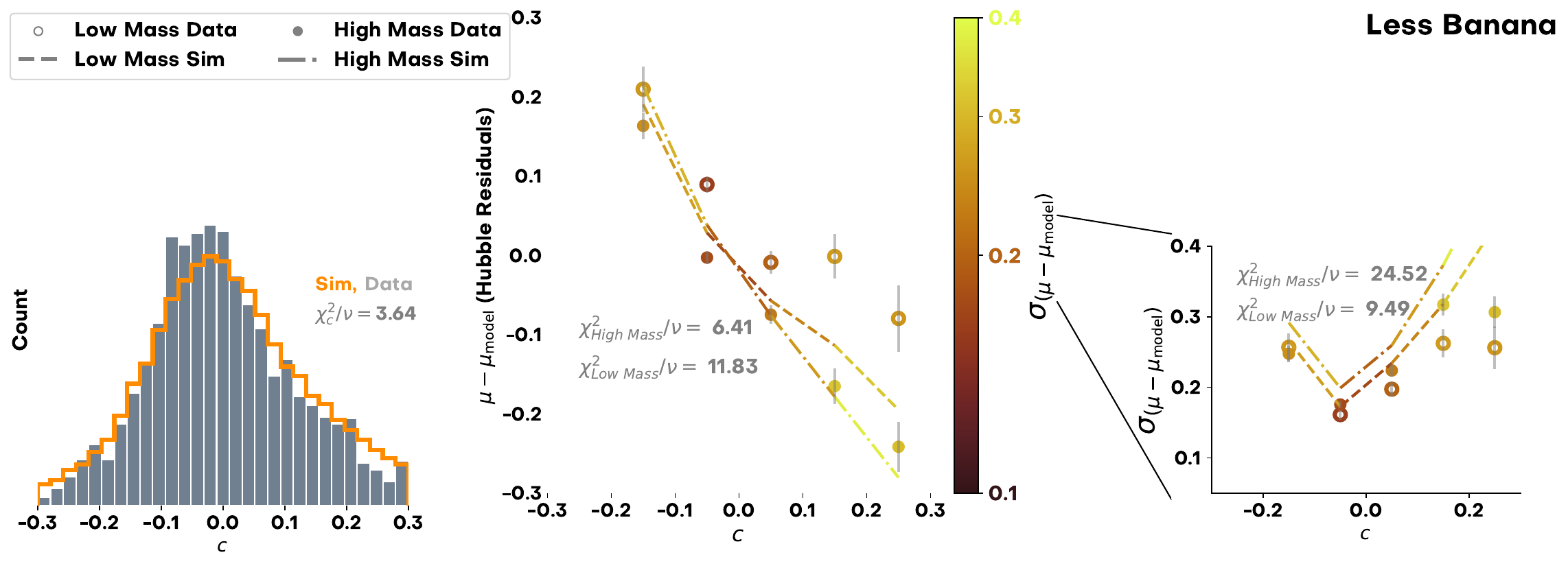}
    \caption{\texttt{Dust2Dust} metrics for the Less Banana Split model. }
    \label{fig:LBN_METRIC}
\end{figure*}

\section{Stjörnumál Dependence on Nuisance Parameters}\label{sec:appendix:sec:nuisance}

Here we test the Tripp nuisance-parameter ($\alpha_{\rm SALT}$, $\beta_{\rm SALT}$) dependence on our inferred parameters with Stjörnumál. For $\beta_{\rm SALT}$, we unsurprisingly find a strong dependence between $\beta_{\rm SALT}$, $\beta_{\rm int}$, and $E(B-V)$, and a weaker dependence on $R_V~\mu$. There is little interaction between the assumed $\alpha_{\rm SALT}$ and our other parameters, though interestingly there does appear to be a weak trend between $\alpha_{\rm SALT}$ and $c_{\rm int}~\sigma$.

\begin{figure}
    \centering
    \includegraphics[width=8cm]{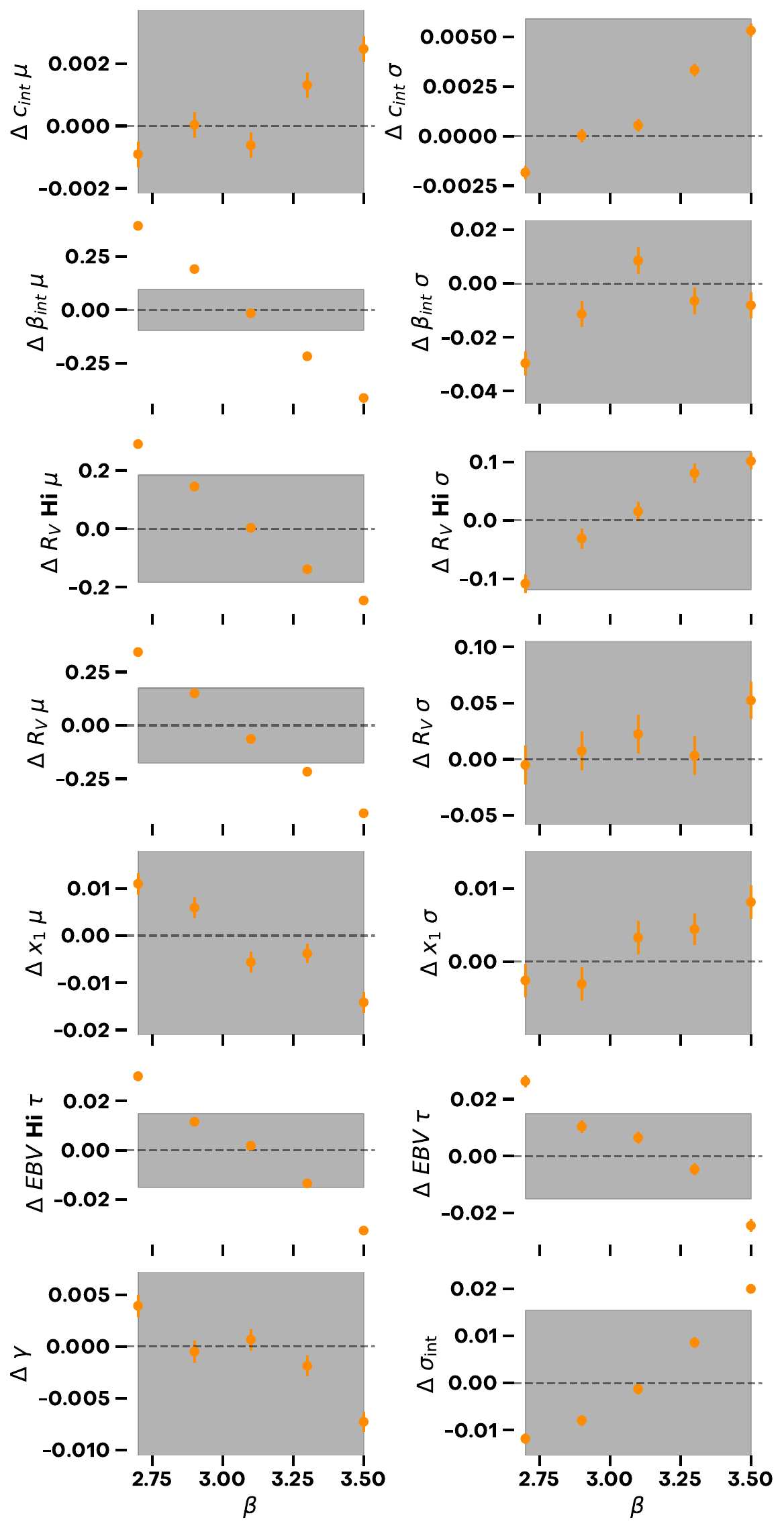}
    \caption{Bias in inferred parameters as a function of $\Delta \beta$ between the model and the simulated cosmology. Orange points show the average bias in the relevant parameter; errors include the $1/\sqrt{N}$ term. Grey fill indicates the $1\sigma$ uncertainty on that parameter as shown in Table \ref{tab:results}. }
    \label{fig:BETA_PARAMS}
\end{figure}

\begin{figure}
    \centering
    \includegraphics[width=8cm]{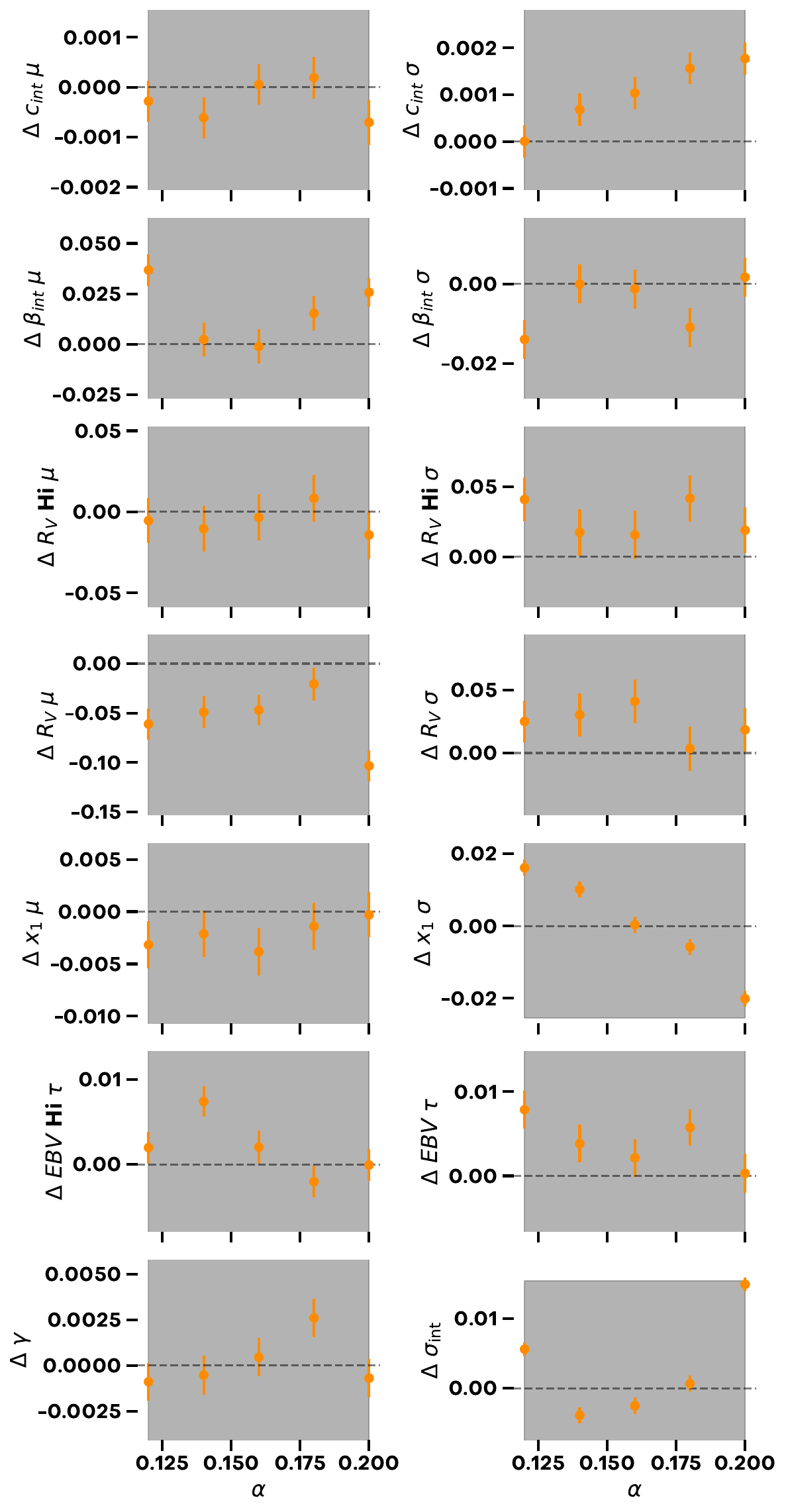}
    \caption{Bias in inferred parameters as a function of $\Delta \alpha$ between the model and the simulated cosmology. Orange points show the average bias in the relevant parameter; errors include the $1/\sqrt{N}$ term. Grey fill indicates the $1\sigma$ uncertainty on that parameter as shown in Table \ref{tab:results}. }
    \label{fig:ALPHA_PARAMS}
\end{figure}



\bibliographystyle{mnras}
\bibliography{research2, matt} 

\bsp	
\label{lastpage}
\end{document}